\documentclass[a4paper,10pt]{article}
\usepackage{graphicx}
\begin{document}

\title{Global stability analysis of axisymmetric boundary layers 
}


\author{Ramesh Bhoraniya         \and
        Narayanan Vinod  
}
 \maketitle
 \begin{abstract}
 This paper presents the linear global stability analysis of the incompressible 
 axisymmetric boundary layer on a circular cylinder.
 The base flow  is parallel to the axis of the cylinder at inlet.
 The pressure gradient is zero in the streamwise direction.
 The base flow  velocity profile is fully non-parallel and 
 non-similar in nature.
 The boundary layer grows continuously in the spatial directions. 
 Linearized Navier-Stokes(LNS) equations are derived for the disturbance 
 flow quantities in the cylindrical polar coordinates.
 The LNS equations along with homogeneous boundary 
 conditions forms a generalized eigenvalues problem.
 Since the base flow is axisymmetric, the disturbances are periodic in azimuthal 
 direction.
 Chebyshev  spectral collocation method and Arnoldi's 
 iterative algorithm is used for the solution of  the  general eigenvalues problem.
 The global temporal modes are computed for the range of Reynolds 
 numbers and different azimuthal wave numbers. 
 The largest imaginary part of the computed eigenmodes are negative and 
 hence the flow is temporally stable.
 The spatial structure of the eigenmodes shows that the flow is 
 convectively unstable because the disturbance amplitudes grow in size 
 and magnitude when moving towards downstream.
 The global modes of axisymmetric boundary layer are more stable than that 
 of 2D flat plate boundary layer at low Reynolds number.
 However, at high Reynolds number they approaches to 2D  flat plate boundary layer.
 Thus, the damping effect of transverse curvature is significant at low Reynolds number.
 The wave-like nature of the disturbances is found for most unstable eigenmodes.
\end{abstract}
\section{Introduction}
The linear stability analysis with the parallel base flow
assumption is termed as local stability
analysis where the base flow is varying only in wall normal direction.
The linear stability of shear flows is governed by classical
Orr-Sommerfeld equation.
%
The numerical solutions to such problems deals with solutions of Orr-Sommerfeld equation
obtained for disturbance amplitude functions are of wave-like nature.
It is assumed that streamwise wavenumber is constant at each streamwise location and
disturbance amplitude functions are a function of wall normal coordinate only.
The numerical solution of Orr-Sommerfeld equation 
parallel base flow assumption can be achieved with very limited 
computational efforts \cite{Drazin}. 
The numerical solutions of Orr-Sommerfeld equation
obtained for disturbance amplitude functions are of wave-like nature.
It is assumed that streamwise wavenumber is constant at each streamwise location and
disturbance amplitude functions are a function of wall normal coordinate only.
In spatially growing boundary layers, the base flow velocity is
varying in streamwise direction which results in  a considerable wall normal velocity.
In these cases, non-parallel effects are very strong and cannot be neglected.
This is true at low and moderate Reynolds numbers.
%
The parallel flow assumption is not a good approximation for real flows where the
flow is varying in streamwise direction. Under such conditions the flow becomes
two-dimensional, and only wave-like  nature of the solution is not expected.
However, it may be one of the solution.
Therefore, stability analysis with the parallel flow assumption may not
represent the complete flow physics,
and hence, it is having very limited scope for parallel and weakly
non-parallel (WNP) flows only.\\

%
In the last decade, the stability analysis of non-parallel flow have received much 
attention through global stability analysis. 
Here the base flow is varying in both, radial and axial directions. 
The term  global instability is used to represent instability analysis 
of the  such flows \cite{Theofilis03}.
The base flow considered here is non-parallel due to continuous 
growth of boundary layer in spatial directions and not self-similar 
due to transverse curvature effect \cite{gl,vinodthesis}.\\
Global temporal modes are time harmonic solutions of the homogeneous linearized
disturbance equations with homogeneous boundary conditions in space. 
Such solution obtained numerically for a base flow with strong non parallel effects
\cite{Zebib}. If the base flow is varying slowly on the scale of a typical instability 
wavelength, global modes can be recovered by WKBJ type analysis \cite{Crighton}.
A boundary layer at high Reynolds numbers and flow through diverging channels at 
small angles of divergence are common examples of weakly non-parallel flows.
The boundary layer at low and moderate Reynolds number have strong non-parallel effects
and hence WKBJ approach may not be the right choice.
The prediction by non-parallel theory also agrees better with the experiments at
moderate Reynolds numbers.\\
%
Joseph is the first among all to introduce the term global stability analysis,
where he studied perturbation energy and established lower bounds for flow 
stability \cite {Joseph}. 
Bert reported the discovery of short-wavelength elliptic
instability in inviscid vortex flows, a problem related 
to both transition and turbulence research.
Bert discovered short-wavelength instability for inviscid vortex 
flow related to transition and turbulence research \cite {bert}. 
Jackson reports first computations on viscous flow where he studied 
flow past various shaped bodies \cite{Jackson}.
onset of periodic behavior in two-dimensional laminar flow past bodies 
of various shapes is examined by means of finite-element simulations.
Zebib studied growth of symmetric and asymmetric perturbations for 
flow past a circular cylinder  \cite{Zebib}.
They found that the symmetric disturbances are stable, 
the asymmetric perturbations become unstable at a Reynolds 
number about 40 with a Strouhal number about 0.12. 
The global approach is used by Christodoulou for free surface flow \cite{christodoulou}, 
Tatsumi for rectangular duct flow \cite{Tatsumi} and Lin for boundary layer flows \cite{Lin}.
The matrices generated by the discretization of global stability equations are very large, 
non-symmetric and sparse, which requires significant computational resources.
All above said work have implemented QR algorithm for computations of eigenvalues
and eigenvectors of resulting matrices.
This algorithm computes all the eigenvalues. 
However, flow becomes unstable in case of a shear flow due to very few dangerous modes only, 
it is more economical to calculate that modes only. 
Iterative techniques like Arnoldi's algorithm with shift-invert strategy computes 
few selected eigenvalues near the shift value.
Theofilis studied 2D steady laminar separation bubble using WNP and DNS and got 
an excellent agreement of the stability results for both 2D and 3D \cite{Theofilis02}. 
By performing a global stability analysis, they proved the existence of new 
instability modes that are not explored by either of the approaches i.e. WNP \& DNS.
Theofilis studied the global stability of separated profiles in three 
different flow configurations, and they show that the amplitude of the global mode is 
less in the separated region than in the wake region or shear layer region \cite{Theofilis03} .
Ehrenstein studied the global instability of a flat plate boundary layer \cite{Ehrenstein} . 
By an optimal superposition of the global temporal modes, they were able to simulate
the convective nature of instability of the boundary layer. 
Alizard performed a global stability analysis on a flat plate boundary layer and obtained 
an excellent agreement with the local stability results and the WNP results \cite{Alizard} . 
Monokrousos performed a global stability analysis of the flat plate boundary layer 
and calculated the maximum energy growth with a reduced order model and found that optimal 
energy growth can not be obtained with least stable global modes only but need to consider 
few globally stable modes also \cite{Monokrousos} .
This is essential for flow control. 
Tezuka has first time solved initial value problem based on the same approach 
for three-dimensional disturbances \cite{Tezuka} . \\

The study of stability analysis of incompressible axisymmetric boundary
layer is found very sparse in literature.
The available literature on axisymmetric boundary layers
are limited to local stability analysis.
Rao first studied the stability of axisymmetric boundary layer \cite{Rao}. 
He found that non-axisymmetric disturbances are less stable than that 
of two-dimensional disturbances. 
The estimated critical Reynolds number based on free stream velocity and 
body radius of the cylinder was 11,000. 
Later work of Tutty investigated that for non-axisymmetric mode critical 
Reynolds number increases with azimuthal wave number N \cite{Tutty}. 
The critical Reynolds number found to be 1060 for N=1 mode and 12439 for N=0 mode. 
The axisymmetric mode is found least stable fourth mode. 
Recently Vinod  investigated that higher non-axisymmetric mode $N=2$ is 
linearly stable for a small range of curvature only \cite{vinodthesis} . 
The helical mode $N=1$ is unstable over a significant length of the cylinder, 
but never unstable for curvature above $1$. 
Thus, transverse curvature has overall stabilizing effect over mean flow and perturbations. 
Malik studied the effect of transverse curvature on the stability of 
incompressible boundary layer \cite{Malik85}. 
They investigated that the body curvature and streamline curvature are having 
significant damping effects on disturbances. 
They also found that traveling waves are the most amplified waves 
in three-dimensional boundary layers.
The secondary instability of an incompressible axisymmetric boundary 
layer is also studied by Vinod  \cite{vinod12}.
They found that laminar flow is always stable at high transverse curvature 
to secondary disturbances.
Parallel base flow assumption is considered in all above investigations. \\
The main aim of this paper is to study the global stability characteristics of 
the axisymmetric boundary layer and the effect of transverse curvature on it. 
The characteristics of the global eigenmodes of axisymmetric boundary layer are 
compared with the flat plate boundary layer(with zero transverse curvature) at 
different Reynolds numbers.The comparison shows that the transverse curvature has
overall stabilizing effect on the disturbances.
\section{ Problem formulation }
We follow the standard procedure  to derive Linearized Navier-Stokes equations 
for disturbance flow quantities. 
The Navier-Stokes equations for mean flow and instantaneous flow are written 
in polar cylindrical coordinates ($r, \theta, x$).
The equations are non-dimensionalised using free stream velocity ($U_\infty$)
and displacement thickness ($\delta{^*}$) at the inlet of the domain.
Following the standard procedure of stability analysis,
we get the set of equations for disturbance flow quantities.
The base flow is axisymmetric  and disturbances
are three-dimensional in nature.
The Reynolds number is defined as,
  \begin{equation}\\
    R_{e}= \frac{U_\infty  \delta{^*}} {\nu}
  \end{equation}
\vskip2mm
The flow quantities are split into  base flow and perturbations,
  \begin{equation}\\
     \overline {U} =U+u,  \\ 
     \overline {V}=V+v , \\ 
     \overline {W}=0+w,  \\
     \overline {P}=P+p , \\
  \end{equation}

We assume normal mode form for disturbances with amplitude varying
in radial($r$) and streamwise($x$) directions. It is correct to assume
periodicity in azimuthal($\theta$) direction as the flow is axisymmetric.

  \begin{equation}\\ 
      q(x, r, t)= \hat q(x, r)e^{[i(N\theta-\omega t)]}
  \end{equation}
             where,\\  
              q = [u, v, w, p]\\
              Q = [U, V, W, P] \\
              $\overline{Q}=[\overline{U},\overline {V},\overline {W},\overline {P}]$\\
              $\omega$ = frequency of the waves \\
              N= azimuthal wave number \\
              $U_\infty$=free stream velocity\\
              $\delta{^*} $ = displacement thickness \\
              $a$ = body radius of the cylinder \\

\noindent    where,  q , Q and $\overline{Q}$ are the perturbation, mean and instantaneous 
    flow quantities respectively.
    u ,v and w are the disturbance velocity components in the axial($x$), radial($r$) and 
    azimuthal($\theta$) directions respectively. 
    The Linearized Navier-Stokes equations for instability analysis are as follows, 
   \begin{equation}
    \frac{\partial u}{\partial t} + U \frac{\partial u}{\partial x}
    + u \frac{\partial U}{\partial x} + V \frac{\partial u}{\partial r}
    + v \frac{\partial U}{\partial r} + \frac{\partial p}{\partial x}
    - \frac {1}{Re} [\bigtriangledown^2 u]=0
     \end{equation}
   \begin{equation}
   \frac{\partial v}{\partial t} + U \frac{\partial v}{\partial x}
    + u \frac{\partial V}{\partial x} + V \frac{\partial v}{\partial r}
    + v \frac{\partial V}{\partial r}+ \frac{\partial p}{\partial r}
    - \frac {1}{Re}[\bigtriangledown^2 v - \frac{v}{r^2}
    - \frac{2} {r^2} \frac{\partial w}{\partial \theta}]=0
  \end{equation}
  \begin{equation}
   \frac{\partial w}{\partial t} + U \frac{\partial w}{\partial x} 
    + V \frac{\partial w}{\partial r} + V \frac{w}{r}
    + \frac {1}{r}\frac{\partial p}{\partial \theta}
    - \frac {1}{Re}[\bigtriangledown^2 w - \frac{w}{r^2} 
    +\frac {2}{r^2} \frac {\partial v}{\partial \theta}]=0
  \end{equation}
  \begin{equation}
   \frac{\partial u}{\partial x} + \frac{\partial v}{\partial r} 
  + \frac{v}{r} + \frac{1}{r} \frac{\partial w}{\partial \theta}=0
  \end{equation}
    where, 
    \begin{equation}
    \bigtriangledown^2 =\frac{\partial^2 }{\partial x^2} 
   + \frac{\partial^2 }{\partial r^2}+\frac{1}{r} \frac{\partial }{\partial r}
   + \frac{1}{r^2} \frac{\partial^2 }{\partial \theta^2}=0
  \end{equation}
   \begin{figure}
     \centerline {\includegraphics[height=1.5in, width=2.5in, 
                 angle=0]{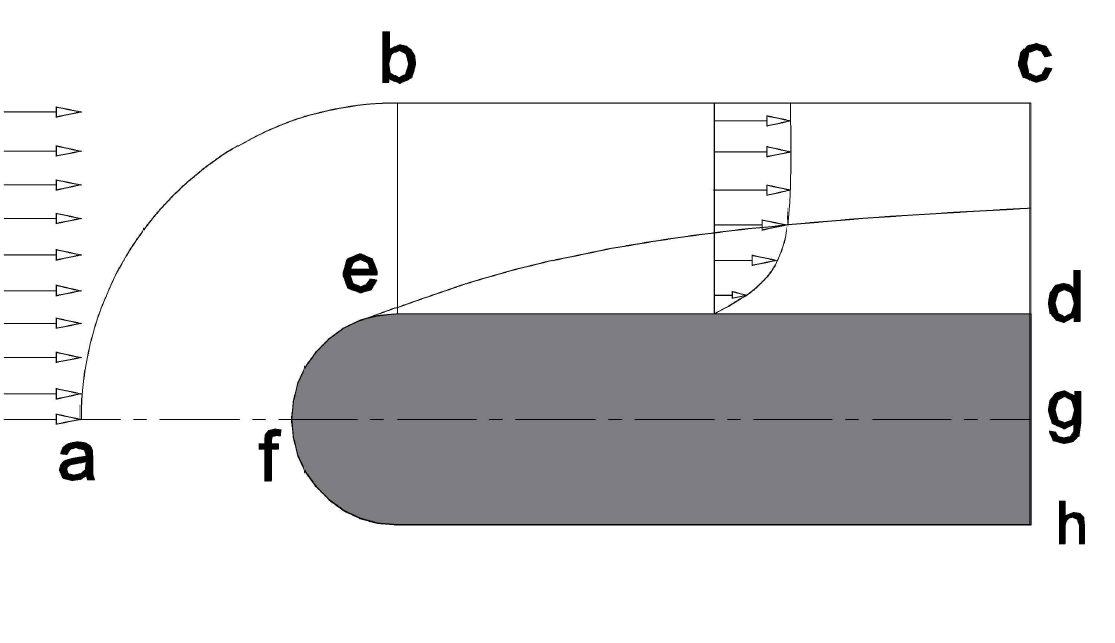}}          
     \caption{Schematic diagram of axisymetirc boundary layer on a circular cylinder.
          Dotted line represents the domain for stability computations. }
   \label{axi_model}
   \end{figure}
   \subsection{Boundary conditions}
    At wall, on the solid surface of cylinder no slip and no penetration 
    boundary conditions are applied.  
    All disturbance velocity components have zero magnitude  
    at the solid surface of the cylinder. \\ 
    \begin{equation} 
      \ u(x, a)=0 ,   \  v(x, a)=0 ,  \  w(x, a)=0 
    \end{equation}  
    At free stream far away from the surface of cylinder exponential 
    decay of disturbances is expected. 
    Homogeneous Dirichlet boundary conditions are applied for velocity and pressure 
    disturbances at free-stream. 
   \begin{equation}
      \  u(x, \infty)=0,  \ v(x, \infty)=0,  \ w(x, \infty)=0 , \ p(x, \infty)=0 
   \end{equation} 
    The boundary conditions in streamwise direction are not straight forward. 
    Homogeneous Dirichlet boundary conditions are applied on disturbance velocity 
    components at the inlet.
    This boundary condition is appropriate as suggested by Theofilis (2003)  
    because we are interested in the disturbances evolved within the basic 
    flow field only.
    At outlet one may apply boundary conditions based on the incoming/outgoing 
    wave information \cite{Fasel}.
    Such conditions impose  wave-like nature of the disturbances and it is
    more restrictive in nature.
    This is not appropriate from the physical point of view in instability analysis.
    Even the wave number $\alpha$ is not known initially in case of a 
    global stability analysis.
    An alternative way is to impose numerical boundary condition which 
    extrapolate information from the interior of the computational domain. 
    Linear extrapolated conditions are applied  by several investigators. 
    Review of literature on global stability analysis suggests
    linearly extrapolated boundary conditions are the good approximations
    \cite{Theofilis03,Swaminathan}.
    We also tried second order extrapolated
    boundary conditions, however the difference is only marginal.
    Thus, we considered linear extrapolated conditions at outlet.
    \begin{equation}
     u(x_{in},r)=0, \ v(x_{in},r)=0, \ w(x_{in},r)=0,
     \end{equation}
     \begin{equation}
     u_{n-2}[x_n-x_{n-1}]-u_{n-1}[x_n-x_{n-2}]+ u_n[x_{n-1}-x_{n-2}]=0 
     \end{equation}
     Similarly, one can write extrapolated boundary conditions for wall normal 
     and azimuthal disturbance components v and w respectively. 
     The boundary conditions for pressure  do not exist physically at the wall.  
    However, compatibility  conditions derived from the Linearized Navier-Stokes 
    equations are collocated at the wall of cylinder \cite{Theofilis03} .
    \begin{equation}\\
    \frac {\partial p}{\partial x}= \frac{1}{Re}[\bigtriangledown^2 u]
    -U\frac{\partial u} { \partial x}- V\frac{ \partial u}{ \partial r}  
    \end{equation}
    \begin{equation}\\
    \frac{\partial p}{\partial r}=\frac{1}{Re}[\bigtriangledown^2 v]
    -U\frac{\partial v}{\partial x}- V\frac {\partial v}{\partial r}
    \end{equation}
     The Linearized Navier-Stokes equations are discretized using 
     Chebyshev spectral collocation method. 
     The Chebyshev polynomial generates non-uniform grids and generates more 
     collocation points towards the ends. 
     It is a favourable arrangement for the boundary value problems.
     \begin{equation}
       x_{cheb}= \cos(\frac{ \pi i}{n})\quad {\rm where} \quad i= 0, 1, 2, 3. . . . n
     \end{equation}
     \begin{equation}
        y_{cheb}= \cos(\frac{ \pi j}{m})\quad {\rm where} \quad j= 0, 1, 2, 3. . . . m
     \end{equation}
      Where n and m are number of collocation points in streamwise and wall normal direction. 
      The gradient of disturbance amplitude functions is very large near 
      the wall region within the thin boundary layer,
      which requires a large number of grid points to increase the spatial resolution. 
      Grid stretching is applied via the following algebraic equation \cite{Malik}.
      \begin{equation}
         y_{real}= \frac{y_i*L_y*(1-y_{cheb})}{L_y+y_{cheb}*(L_y-2y_i)}
      \end{equation}
      In the above grid stretching method, half number of the collocation points 
      are concentrated within the $y_i$ distance from the lower boundary only.
      The non-uniform nature of distribution for collocation points in the 
      streamwise direction is undesirable. 
      The maximum and minimum distance between the grid points are 
      at center and end respectively. 
      Thus, it makes a poor resolution at the center of the domain and a very 
      small distance between the grids at the end gives rise to Gibbs phenomenon. 
      To improve the resolution and to minimize the Gibbs oscillation in the solution, 
      grid mapping is implemented in streamwise direction using following 
      algebraic equation \cite{Costa} .
      \begin{equation}
      x_{map}= \frac {sin^{-1}(\alpha_m x_{cheb})}{sin^{-1}(\alpha_m)}
      \end{equation}
      The value of $\alpha_m$ is selected carefully to improve spatial 
      resolution in the streamwise direction. 
      A very small value of $\alpha_m$ keeps the grid distribution like 
      Chebyshev and near to unity almost uniform grid. 
      For the detail description of the grid mapping readers 
      are suggested to refer \cite{Costa}. 
      To incorporate the effect of physical dimensions of the 
      domain [$L_x$, $L_y$] along with grid stretching and mapping it is 
      required to multiply the Chebyshev differentiation matrices by proper Jacobean matrix. 
      Once all the partial derivatives of the LNS are discretized by spectral 
      collocation method using Chebyshev polynomials, the operator
      of the differential equations formulates the matrices A and B. 
      These matrices are square, real and sparse in nature, 
      formulates general eigenvalues problem.
      \begin{equation}\\
      \left[ \begin{array}{cccc} A_{11} & A_{12} & A_{13} & A_{14} \\ 
         A_{21} & A_{22} & A_{23} & A_{24} \\
         A_{31} & A_{32} & A_{33} & A_{34} \\ A_{41} & A_{42} & A_{43} & A_{44}  
      \end{array} \right]  \left[ \begin{array}{c} u \\ v  
      \\ w \\ p \end{array} \right]= i\omega  
      \left[ \begin{array}{cccc} B_{11} & B_{12} & B_{13} & B_{14} \\ 
         B_{21} & B_{22} & B_{23} & B_{24} \\ B_{31} & B_{32} & B_{33} & B_{34} \\
         B_{41} & B_{42} & B_{43} & B_{44} \end{array}  
      \right]  \left[ \begin{array}{c} u \\ v \\ w \\ p \end{array}
      \right]
      \end{equation} 
      $$
      [A][\phi] = i\omega [B][\phi]     
      $$   
      where A and B are square matrix of size 4nm,  $ i\omega$ is an eigenvalues 
      and $\phi$ is a vector of unknown amplitude of disturbance  
      flow quantities u, v, w and p.  The above mentioned all the boundary 
      conditions are properly incorporated in the matrix A and B. 
\subsection{Solution of general eigenvalues problem}
The matrix A and B are sparse in nature and of very large size. 
For $ n=121$ and $m=121$ the number of eigenvalues are $58564$.
i.e. in the order of $(10^{4} to 10^{5})$. 
The eigenvalues problem being very large to solve for all the eigenvalues. 
However for instability analysis, the few eigenvalues with its largest 
imaginary parts, which makes the flow temporally unstable, are important. 
Hence, we are interested in the few eigenvalues and corresponding eigenvectors only. 
The QZ algorithm computes the full spectrum of eigenvalues and hence 
obviously it is not economical. 
The iterative method based on Arnoldi’s algorithm is a proper choice. 
The Krylov subspace provides the possibility of extracting major part
of the spectrum using shift and invert strategy. 
The computations of Krylov subspace along with Arnoldi's 
algorithm applied to eigenvalues problem becomes easy.
\begin{equation}
      (A-\lambda B)^{-1} B\phi= \mu\phi \quad  
     where \quad \mu=\frac{1}{i\omega-\lambda}
\end{equation}
Where, $ \lambda$ being the shift parameter and $\mu$ 
is the eigenvalues of the converted problem.
Sometimes it is also called spectral transformation, 
which converts generalized eigenvalues problem to standard eigenvalues problem. 
The Krylov subspace may be computed by successive resolution of the linear 
system with matrix $(A-\lambda B)$, using LU decomposition.
Full spectrum method is employed for this small subspace to get a good 
approximate solution to the original general eigenvalues problem \cite{Theofilis11}. 
The large size of the Krylov subspace extracts major part of the spectrum. 
Generally, the computed spectrum is always nearby the shift parameter. 
The good approximation of shift parameter reduces the number of 
iterations to converge the solution to required accuracy level. 
However, larger size of subspace makes the solution almost 
independent from the shift parameter $\lambda$. 
We tested the code for several values of shift parameter $\lambda$.
The convergence of the solution depends on the value of shift parameter. 
The good approximation of the shift value needs less number of iterations. 
However we have taken the maximum number of iterations equal to 300; 
hence convergence of the solution may not be affected by shift parameter. 
Given the large subspace size of $k=250$, the part of the spectrum for our
instability analysis could be recovered in the one computation that took
about 2.36 hours on Intel Xenon(R) CPU E5 26500@$2. 00GHz\times 18$. 

\section{Base flow solution}
The base flow velocity profile is obtained by solving the steady 
Navier-Stokes equations (21)-(23)with the finite-volume code ANSYS Fluent.
The axisymmetric domain is considered with 20 and 200 
radius in radial and axial directions as shown in figure \ref{axi_model}. 
%
\begin{equation}
   U \frac{ \partial U } { \partial x} + V \frac{ \partial U } { \partial r}= 
- \frac{\partial P}{\partial x}+\frac{1}{R_{e}} (\frac{\partial^2 U}{ \partial x^2} 
+ \frac{1}{r} \frac{ \partial U}{ \partial r}+ \frac{\partial^2 U}{ \partial r^2} )\\
\end{equation}
\begin{equation}
 U \frac{ \partial V } { \partial x} + V \frac{ \partial V } { \partial r}= 
 - \frac{\partial P}{\partial r}+\frac{1}{R_{e}} (\frac{\partial^2 V}{ \partial x^2} 
 + \frac{1}{r} \frac{ \partial V}{ \partial r}+ \frac{\partial^2 V}{ \partial r^2} )\\
\end{equation}
\begin{equation}
  \frac{\partial U}{\partial x} + \frac{\partial V}{\partial r} + \frac{V}{r}=0
\end{equation}
Appropriate boundary conditions are applied to close the formulation of above problem.
Uniform streamwise velocity $U_{\infty}$ was imposed at the inlet (ab) and no-slip
and no penetration boundary conditions at the surface of the cylinder (fed). 
Slip boundary condition($\textbf{n}.\textbf{U}=0$) was imposed at free stream far away 
from the solid surface (bc) and outflow boundary conditions at outlet(cd), that consists $ 
\frac{\partial U} {\partial x}=0 $, $ \frac{\partial V}{\partial x}=0$ and $ P=0 $. 
Line afg is the axis of the cylinder.
The steady Navier-Stokes equations was solved using SIMPLE algorithm with 
under-relaxation , to get a stable solution .
Quick, a weighted average of second order upwind and second order central
scheme was used for spatial discretization of momentum equations. 
Thus obtained base flow velocity profile is interpolated to spectral grids 
using cubic spline interpolation to perform global stability analysis.  
The solution for the base flow has converged with the total 251251 number of grid points 
with 1001 and 251 grids in streamwise and radial directions respectively. 
The distribution of the grids is geometric in both the directions. 
We started simulation initially with  coarse grid size of 501 and 125 grids 
in axial and radial direction and refined it with a factor of 1.4142 in each directions. 
The discretization error was computed through Grid Convergence Index study 
on three consecutive refined grids\cite{Roache}.
The monotonic convergence is found for all this refined grids. 
In our problem we computed error and GCI for two field values 
U(x=75,r=1.2) and V(x=75,r=1.2) near 
the surface of the cylinder, where the gradient of the velocity field 
is maximum. 
The GCI and error was computed for three different grids as shown table \ref{mean_gci} 
The error between Mesh 1 and Mesh 2 is too small. 
The GCI has also reduced with the increased spatial resolution.
It shows that solution is converging monotonically towards the grid-independent one. 
Further refinement in the grid will hardly improve the accuracy of the solution 
while increases the time for the computations. The grid is thoroughly refined near the wall
where velocity gradient is very high. The convergence order n for U and V is 2.00, 
which is in agreement with the second order discretization scheme used in the 
finite volume code ANSYS FLUENT.
A Mesh 1 was checked with domain height of 25 times cylinder radius 
to study the effect of far-field condition.
The percentage error with this two domain height is within 0.1 \%. 
Thus, Mesh 1 was used in all the results presented here in to 
compute velocity field for the base flow.

\begin{figure}       
\includegraphics[height=1.25in, width=2.2in, angle=0]{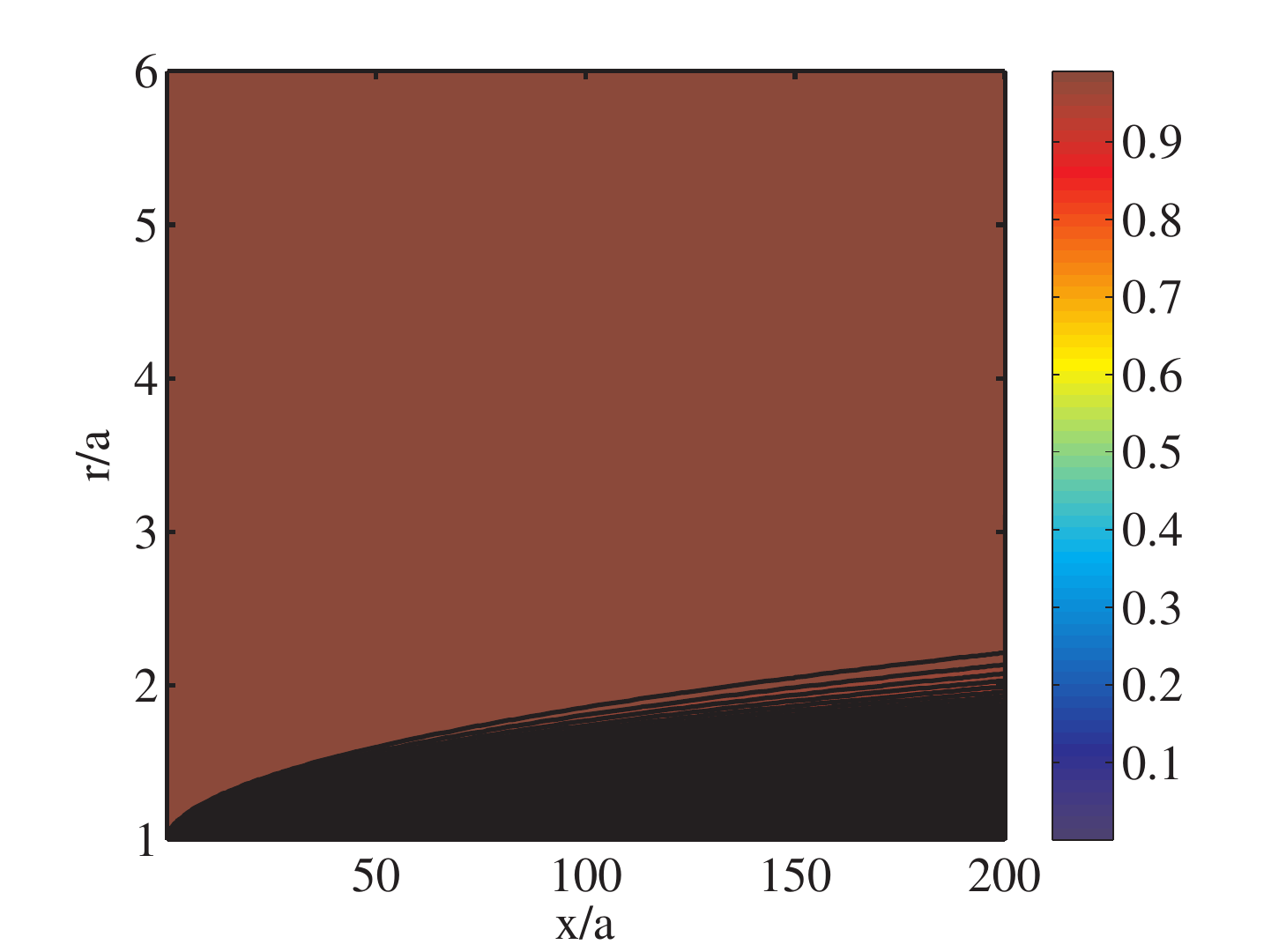}   
\includegraphics[height=1.25in, width=2.2in, angle=0]{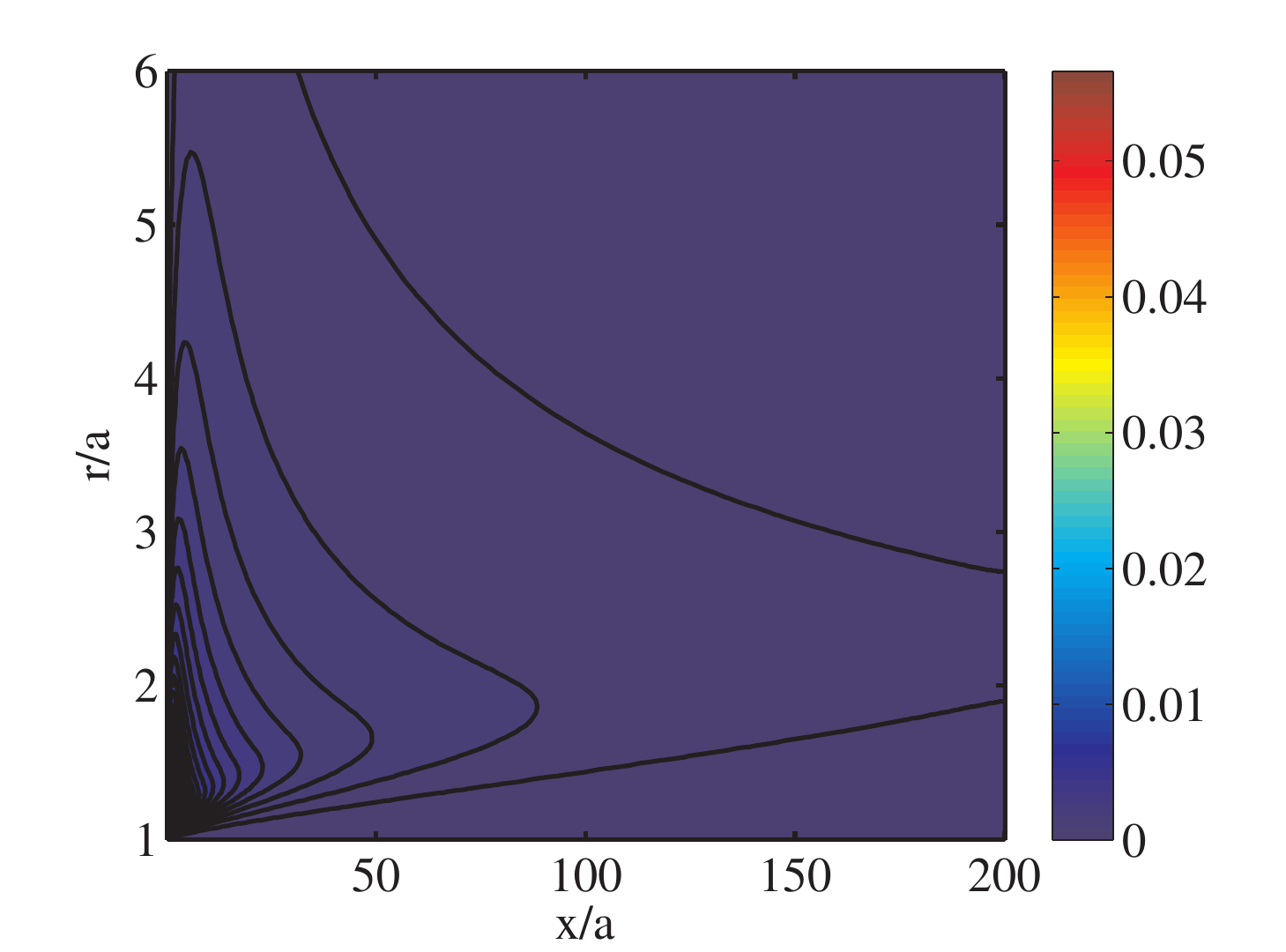}
\caption{Base flow velocity profile for streamwise (U) and  normal(V) velocity 
 components for $R_{e}=2000$. The Reynolds number is calculated based on body 
 radius of the cylinder. The velocity profile is interpolated to collocation 
 points for instability analysis. The actual domain height in wall normal 
 direction is 20 radius of the cylinder. }  
\end{figure}
\begin{figure}       
\includegraphics[height=1.25in, width=2.0in, angle=0]{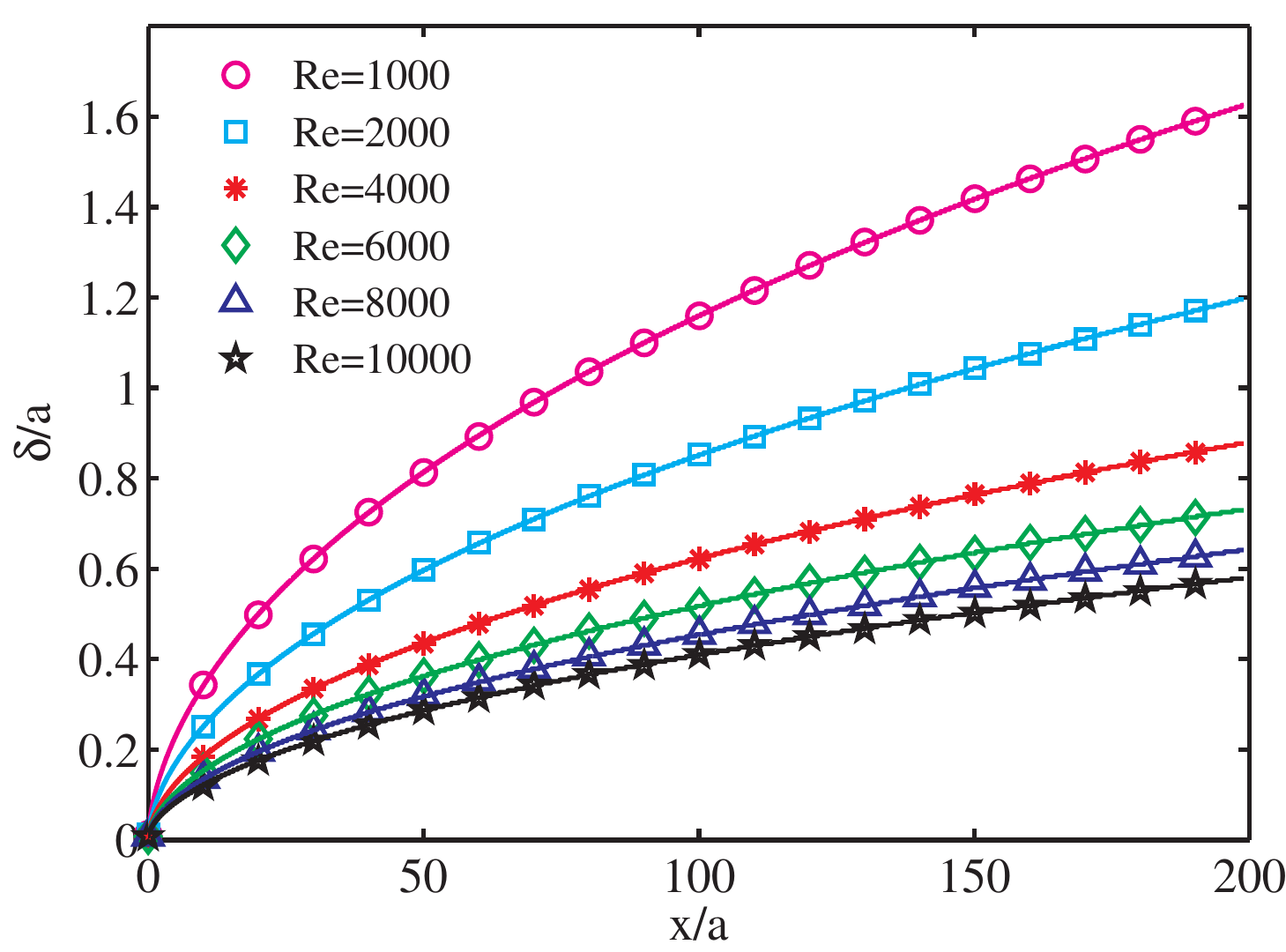}   
\hspace{5px}
\includegraphics[height=1.25in, width=2.0in, angle=0]{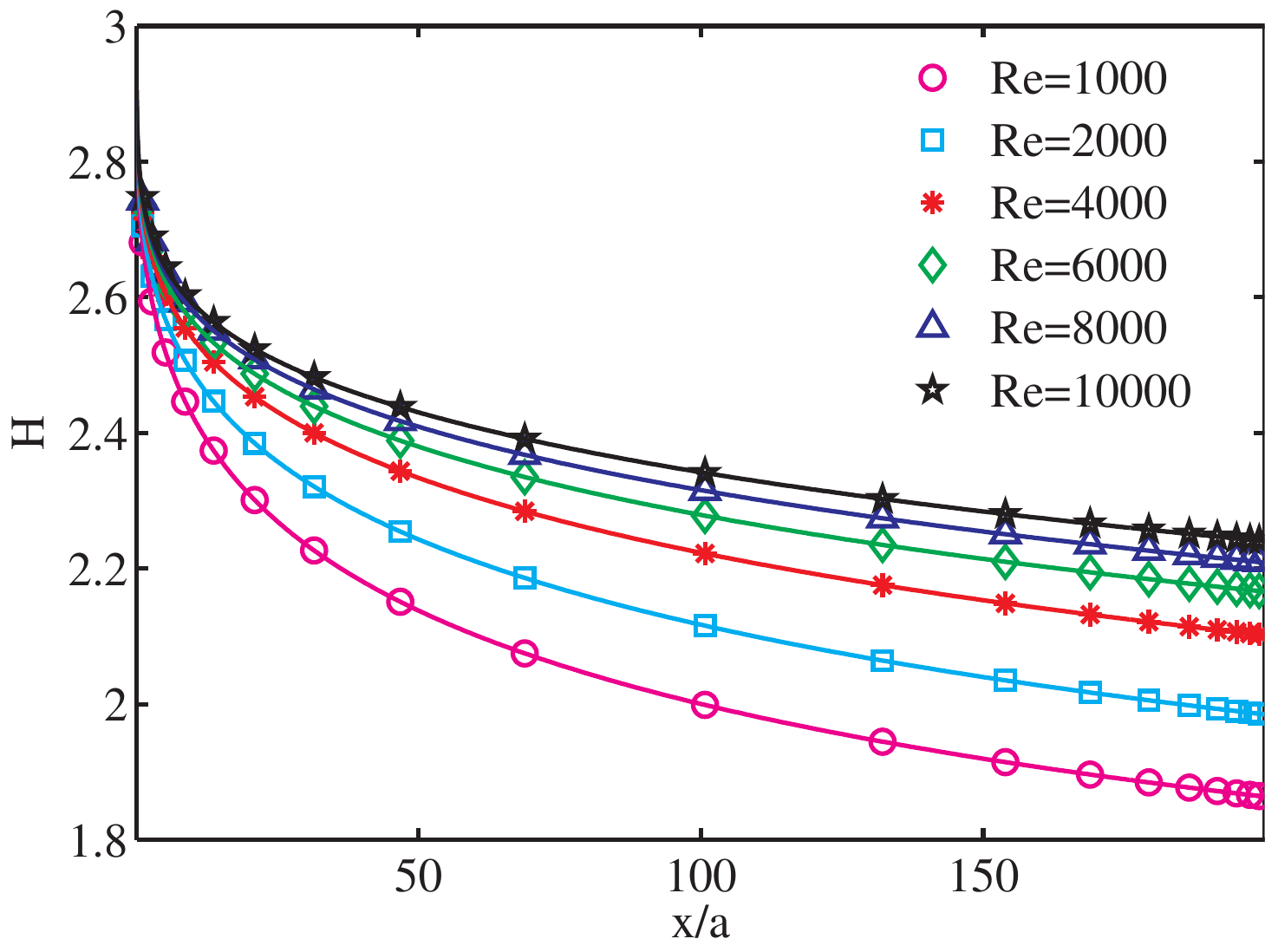}
\caption{ Variation of transverse curvature ($\delta$/a)(left) and shape 
factor(H)(right) in streamwise direction for different Reynolds number. 
Reynolds number is based on body radius(a) of the cylinder. Axial and radial length is 
normalized with cylinder radius (a). }  
\label{curvature}
\end{figure}
\begin{table}
\caption {The Grid Convergence study for the base flow is obtained , 
         using U(x=75,r=1.2) and V(x=75,r=1.2) for Re=1000. 
         The grid refinement ratio($\alpha$)in each direction is 1.4142.
         The relative error($\epsilon$) and Grid Convergence Index (GCI) 
         are calculated using two consecutive grid size. 
         The j and j+1 represents course and fine grids respectively.
         $\epsilon$=$\frac{f_{j}-f_{j+1}} {f_{j}}$ $\times$ 100. 
         $GCI(\%)$=$3[\frac{f_j-f_{j+1}}{f_{j+1}(\alpha^n-1)}]$ $\times$ 100, 
         where n=log[$\frac{f_{j}-f_{J+1}}{f_{j+1}-f_{j+2}}$]/log({$\alpha$}) .}                                                           
\label{mean_gci} 
\begin{tabular}{llllllll}
\hline\noalign{\smallskip}
 Mesh  & Grid size  & U & $\epsilon(\%)$ & GCI(\%) & V & $\epsilon (\%)$&GCI(\%) \\
\noalign{\smallskip}\hline\noalign{\smallskip}
 \#1 & 1001 $\times$ 251 & 0.034138 & 0.00879 & 0.0262& 1.655$\times$ $10^{-5}$ & 0.060&0.1812 \\
 \#2 & 708 $\times$  177 & 0.034135 & 0.0175 & 0.05274 & 1.656$\times$ $10^{-5}$ & 0.12&0.3622 \\
 \#3 & 501  $\times$ 125 & 0.034129 & -- & -- & 1.658$\times$ $10^{-5}$ & --& -- \\
\noalign{\smallskip}\hline
\end{tabular}
\end{table}
 \section{Code validation}
 For the purpose of preliminary comparisons of global stability 
 results with local stability results, we reduced the global nature 
 of the problem to an equivalent local stability problem. 
 The domain length equal to one wavelength (Lx=2$\pi$/$\alpha$) is taken
 in streamwise direction for a least stable axisymmetric mode(N=0). 
 Non-parallel effects from the mean flow velocity profile are removed to 
 apply parallel flow assumption i. e. $ V=0 $ \& $ \frac{ \partial U}{\partial x}=0$. 
 In wall normal direction, the boundary conditions are same as
 that of local stability analysis  \cite{Theofilis03}. 
 To impose wave-like behaviour of disturbances, Robin and periodic 
 boundary conditions are applied in the streamwise direction at 
 inlet and outlet \cite{Swaminathan} . 
 The Robin boundary conditions with constant streamwise wavenumber
 $\alpha$ is prescribed at inlet and outlet. 
 The Robin boundary conditions are derived from the 
 $\phi(r,t)=\hat \phi(r)e^{[(i(N\theta-\omega t)]}$.  
 We applied Robin boundary conditions along with periodic boundary 
 conditions at inlet and outlet with constant streamwise wavenumber $\alpha$.
  \begin{equation}
      \frac{\partial^2 u}{\partial x^2}=-\alpha^2 u \\,  
      \frac{\partial^2 v}{\partial x^2}=-\alpha^2 v \\ 
  \end{equation}
  \begin{equation}
      u(x,r)=u(x+Lx,r)\\, v(x,r)=v(x+Lx,r)\\
  \end{equation}
 
 \begin{figure}
 \centerline {\includegraphics[height=2.0in, width=3.5in, angle=0] 
             {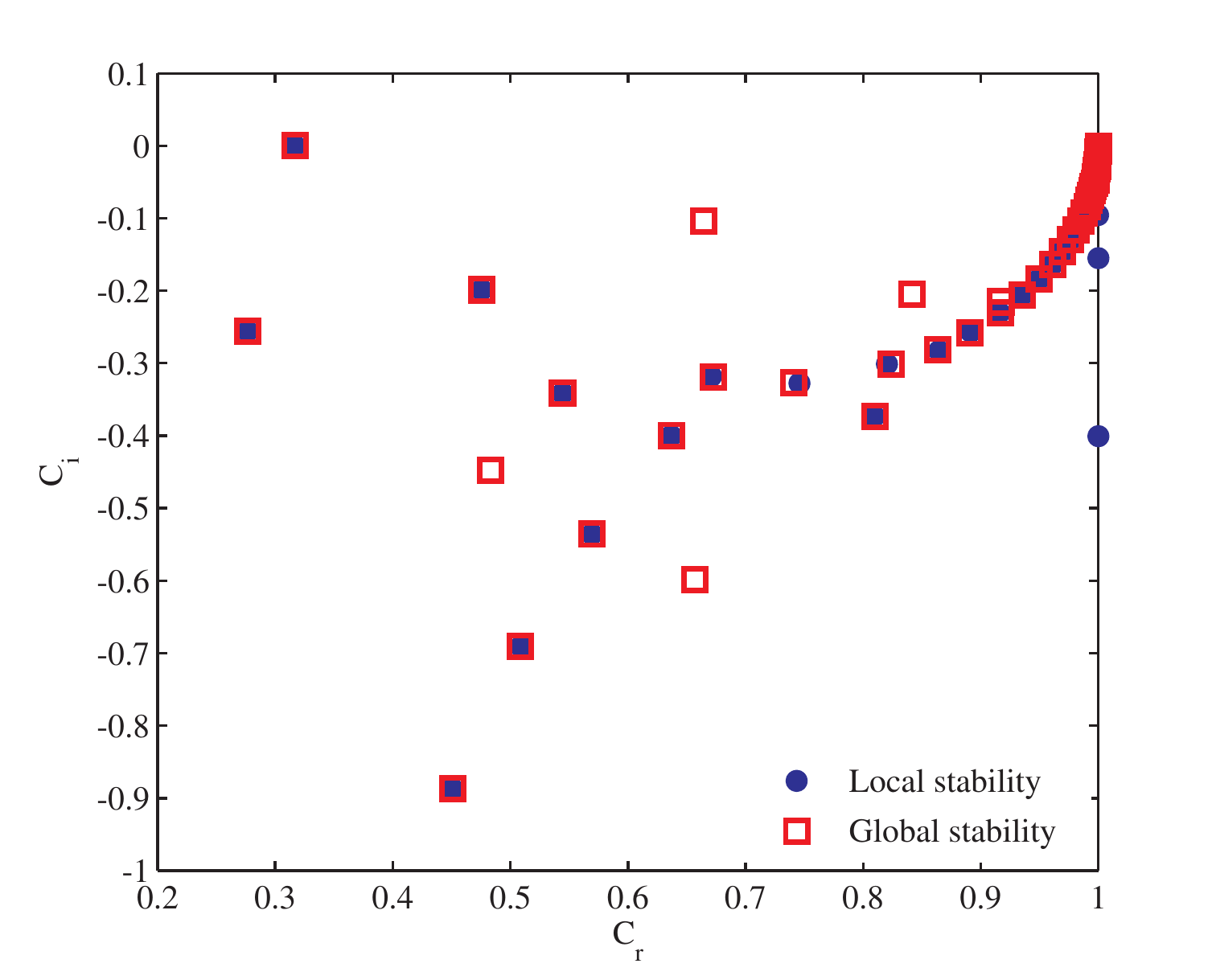}}          
 \caption{Comparison of eigenspectrum for local and global stability analysis 
          for axisymmetric mode (N=0) and Reynolds number $R_{e}=12439$.  
          Here $R_{e}=12439$ is a critical Reynolds number 
          for local stability analysis based on body radius of the cylinder. }
 \label{spectrum}
 \end{figure}
 \begin{figure}      
 \includegraphics [height=2.5in, width=2.5in, angle=0]{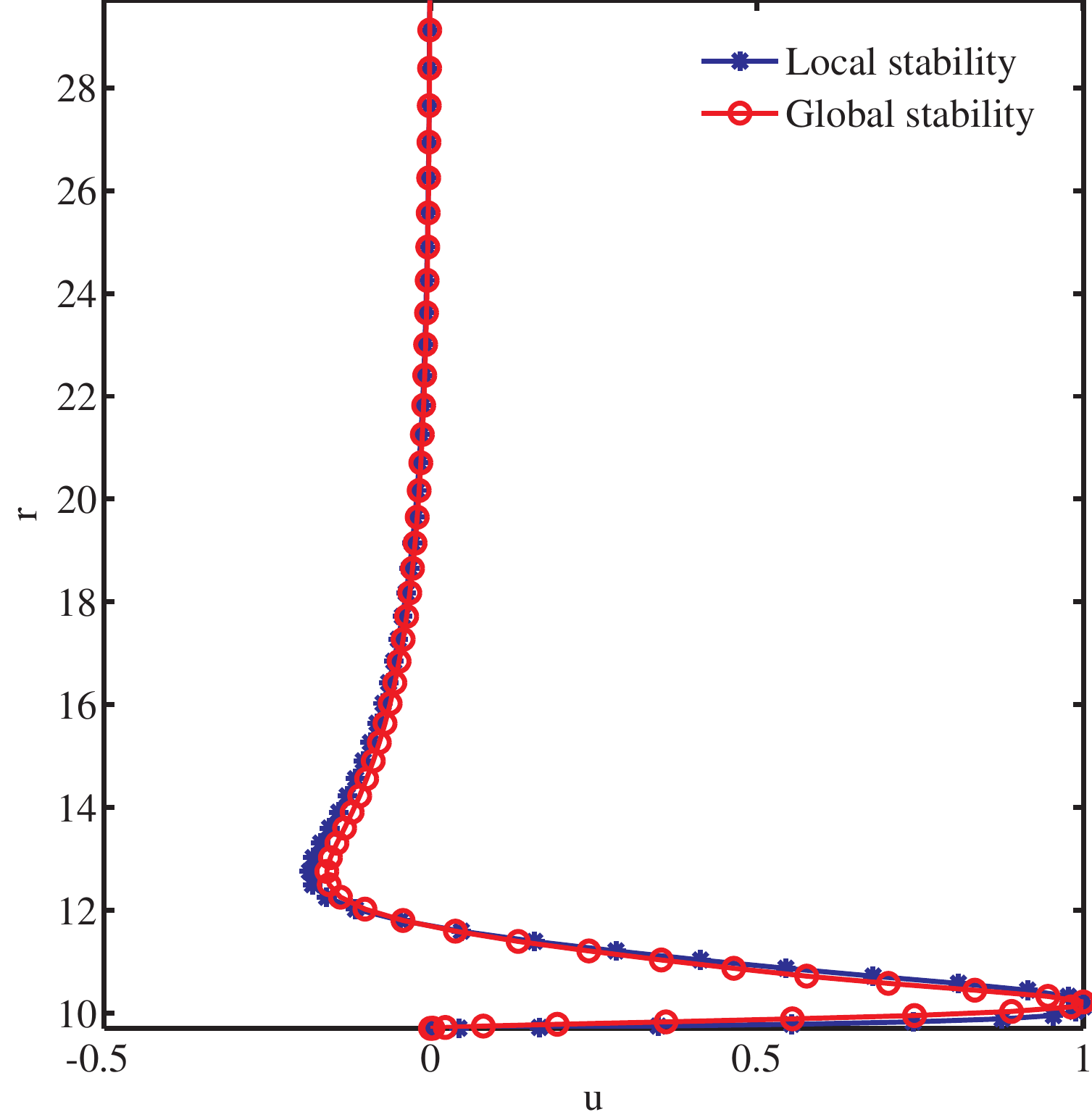}   
 \includegraphics [height=2.5in, width=2.5in, angle=0]{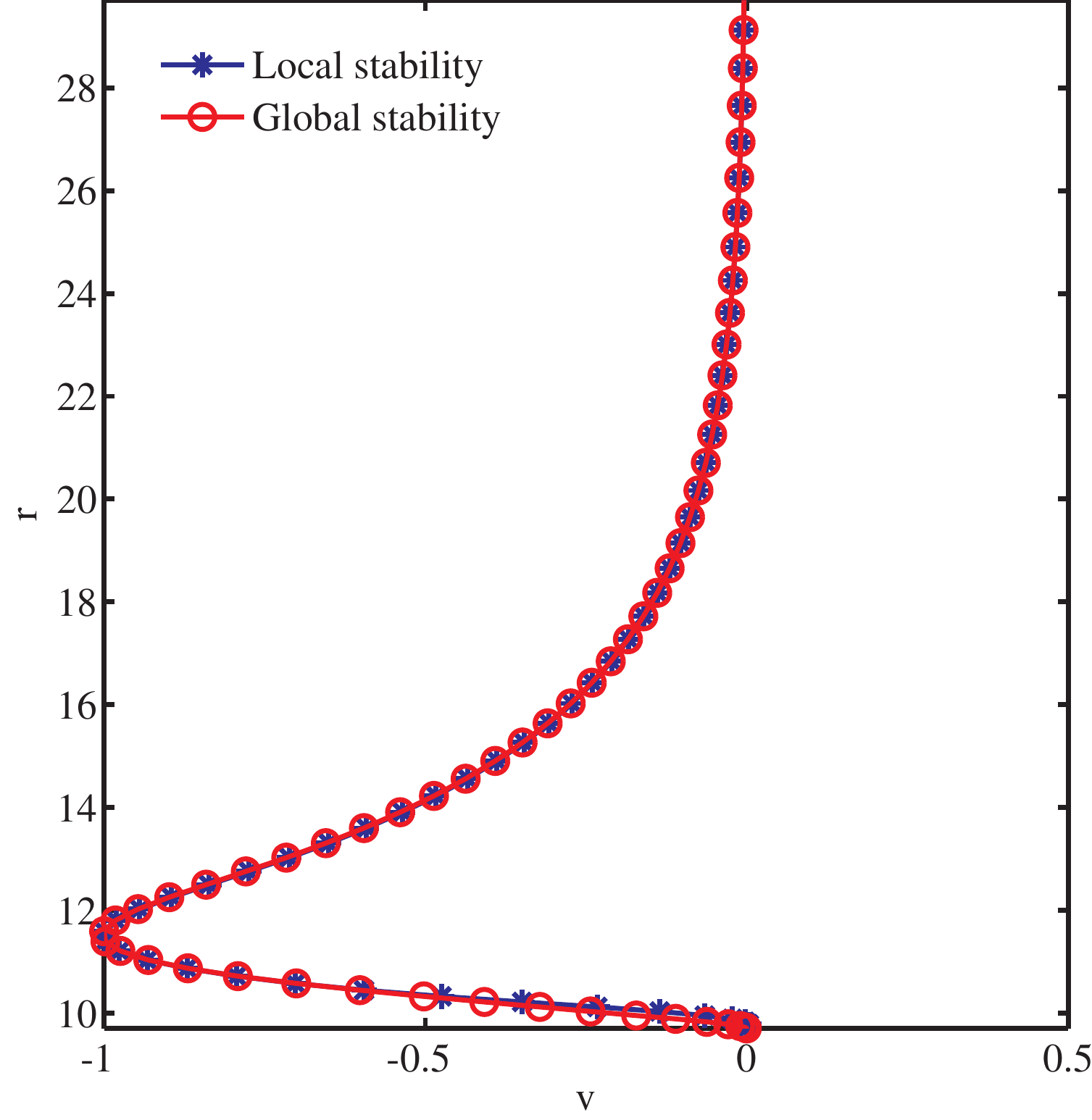}
 \caption{Comparisons of the real parts of (a) streamwise u and 
         (b) wall normal v  eigen functions for local 
         and global stability analysis for axisymmetric mode 
         (N=0) for Reynolds number $R_{e}=12439$ based on body radius of the cylinder. 
         The Reynolds number based on the displacement thickness ($\delta{^*}$) is 1264. }
 \label{eig_fun}
 \end{figure}
 Above boundary conditions are applied to disturbance velocity components 
 in a streamwise direction at inlet and outlet. 
 The second derivatives are considered to avoid complex quantities in 
 the boundary conditions and hence fast computations can take place. 
 Here we considered axisymmetric mode($N=0$) for the validation
 of global stability results. 
 The critical Reynolds number, streamwise wavenumber and streamwise location  
 are 12439, 2.73 and 47 respectively \cite{Tutty}.
 The Reynold number is based on the body radius of the cylinder. 
 Streamwise domain length is taken equal to one wavelength($L_x=2.3015$). 
 The domain in streamwise direction is very limited and hence the solution
 is converged with the number of collocation points n=41 in streamwise 
 direction and m=101 in the radial direction. 
 The eigenspectrum and eigenfunctions are in good agreement with 
 the results of local stability analysis as shown in figure 
 \ref{spectrum} and figure \ref{eig_fun}. 
 The streamwise wavenumber $\alpha $ is computed from the 
 eigenfunctions obtained through global stability analysis
 and it is equal to 2.73. Hence, the applied value of $\alpha$ 
 is recovered in the global stability analysis. 
 The perturbation for the disturbances u and v can be written as 
 $ u=u e^{ [i\theta_u(x, r)]}$ and , $v=v e^{[i\theta_v(x, r)]}$, 
 where $\theta_u $ and $ \theta_v $ are phase angles of wave with 
 respect to x and r. 
 The phase angles of waves can be defined by the following relationship \cite{Alizard}. 
 \begin{equation}
 \theta_u=tan^{-1}(\frac{u_i}{u_r}),    \theta_v=tan^{-1}(\frac{v_i}{v_r}) \\
 \end{equation}
 Where r and i denotes real and imaginary parts of the eigenfunctions u and v. 
 The phase is dependent on the normal coordinate only.  
 Thus the position of maximum absolute magnitude of each disturbances remains 
 same at all the streamwise locations.  
 The streamwise wave number ($\alpha$) is calculated by \cite{Alizard}.            
 \begin{equation}
 \alpha_{ru}= \frac{\partial \theta_u(x, y_{umax)}} {\partial x}, 
 \alpha_{rv}=\frac{\partial \theta_v(x, y_{vmax)}}{\partial x} 
 \end{equation}
 \section{Results and discussions}
 In the present analysis Reynolds number is varying from 261 to 693 with 
 azimuthal wave numbers $0$, $1$, $2$, $3$, $4$ and $5$. 
 Reynolds number based on the displacement thickness ($\delta^*$) 
 at the inlet of the domain is considered.
 Streamwise and wall normal dimensions are normalized with the displacement  
 thickness at the inlet of the domain.
 The streamwise domain length $L_{x}=345$ and wall normal height $L_{r}=20$
 is selected for the instability analysis. 
 The number of collocation points considered are $n=121$ in streamwise direction 
 and m=121 in wall normal direction. 
 The general eigenvalues problem is solved using ARPACK, 
 which uses Arnoldi's algorithm. 
 The computed eigenvalues are found to be accurate up to three decimal points. 
 The additional lower resolution cases were also run to confirm the proper 
 convergence of the eigenspectrum. 
 Heavy sponging is applied to avoid the spurious reflections at the outlet. 
 The eigenvalues with the largest imaginary part that is least stable ones is selected. 
 The two-dimensional mode structure of the selected modes is also checked for the spurious mode.\\ 
 \subsection {Grid convergence study}
\begin{figure}
 \centerline {\includegraphics[height=1.75in, width=2.5in, angle=0]
                              {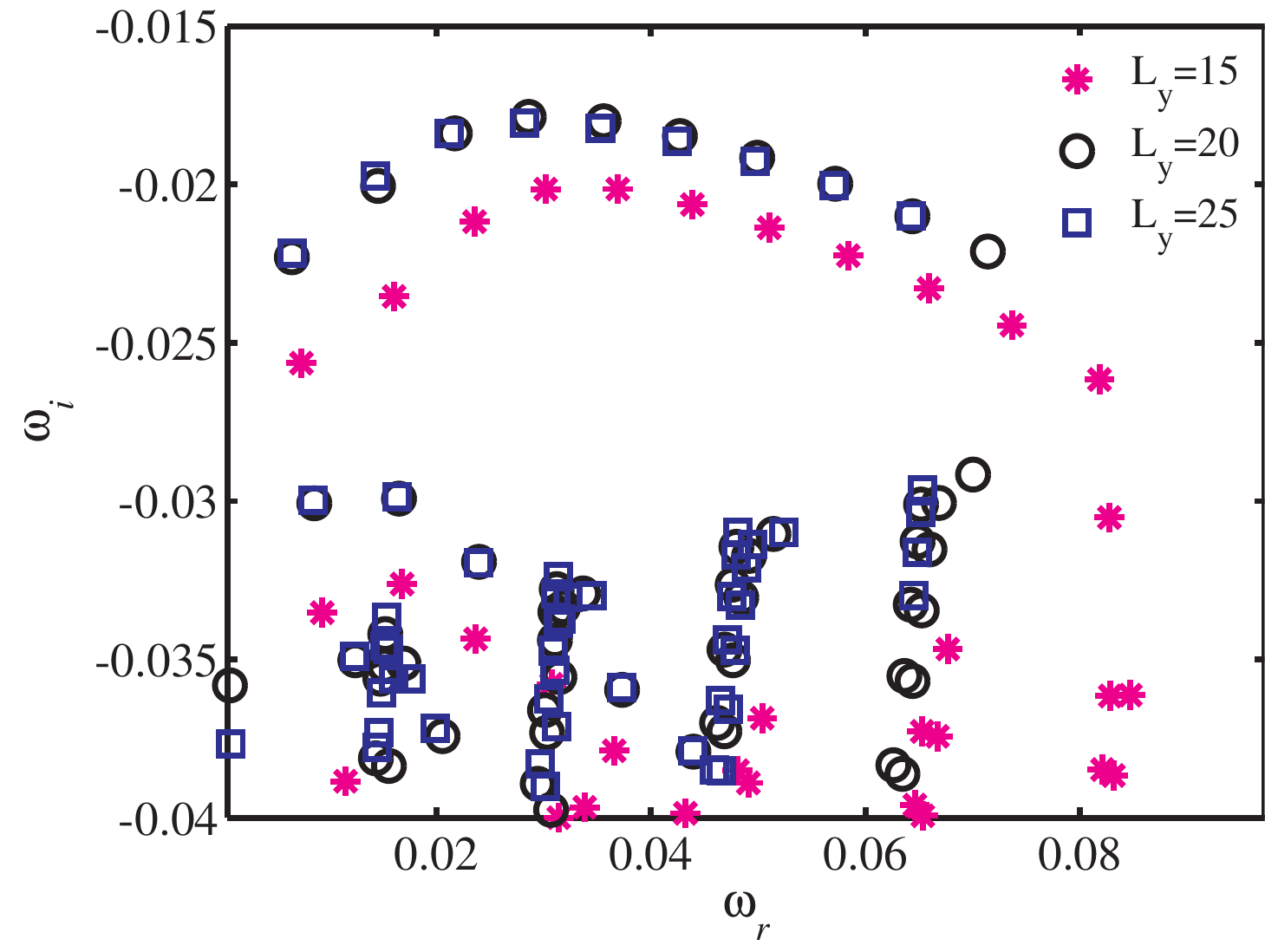}}          
 \caption{Eigenspectrum for azimuthal wavenumber N=1 and Re=383 for 
         three different domain height Lr=15, 20 and 25.}
 \label{sp_Ly}
 \end{figure}

 A grid convergence study was performed to check the accuracy of the 
 solution and appropriate grid size.
 Table 2 shows the values of two leading eigenvalues computed 
 for $Re=383$ for axisymmetric mode using different three grid size.
 The grid resolution was successively improved by a factor of 1.14 
 in axial and radial direction respectively.
 The real and imaginary parts of the eigenvalues shows monotonic convergence 
 of the solution with the increase in resolution. 
 In this table n and m indicates the number of grids in axial and 
 radial directions respectively.
 The relative errors are calculated between consecutive grids for real 
 and imaginary parts of the eigenvalues.
 The largest associated error among both the eigenmode is considered.
 The relative error for mesh 1 is well within the limit and it is used 
 for all the results reported here. 
 \begin{table}
 \caption{The Grid Convergence study for two leading eigenvalues $\omega_{1}$ 
         and $\omega_{2}$ for $R_{e}=383$ and N=1 for different grid size. 
         The grid refinement ratio in each direction is 1.14.
         The maximum relative error is shown here . }
 \label{tab:1}       
 \begin{tabular}{lllllllll}
 \hline\noalign{\smallskip}
 Mesh & Lx &Lr & n$\times$m & n & m & $\omega_{1}$ & $\omega_{2}$ & error(\%) \\
 \noalign{\smallskip}\hline\noalign{\smallskip}
 \#1 & 605&20 & 14641 & 121 & 121 & 0.03654-0.01763i & 0.02965-0.01770i & 3.1598 \\
 \#2 & 605&20 & 11449 & 107 & 107 & 0.03644-0.01709i & 0.02951-0.01751i & 6.3791 \\
 \#3 & 605&20 & 8649  & 93  & 93  & 0.03636-0.01642i & 0.02938-0.01646i &   ---  \\
 \noalign{\smallskip}\hline
 \end{tabular}
 \end{table}
 To check the influence of domain height(Lr), computations were  performed with three 
 different domain height of Lr=15,20 and 25. The difference of results obtained with 
 Lr=20 and Lr=25 is too small as shown in figure \ref{sp_Ly}. 
 Thus Lr=20 is selected for all global stability computations.\\
\begin{figure}
\centerline {\includegraphics[height=1.75in, width=2.5in, angle=0] {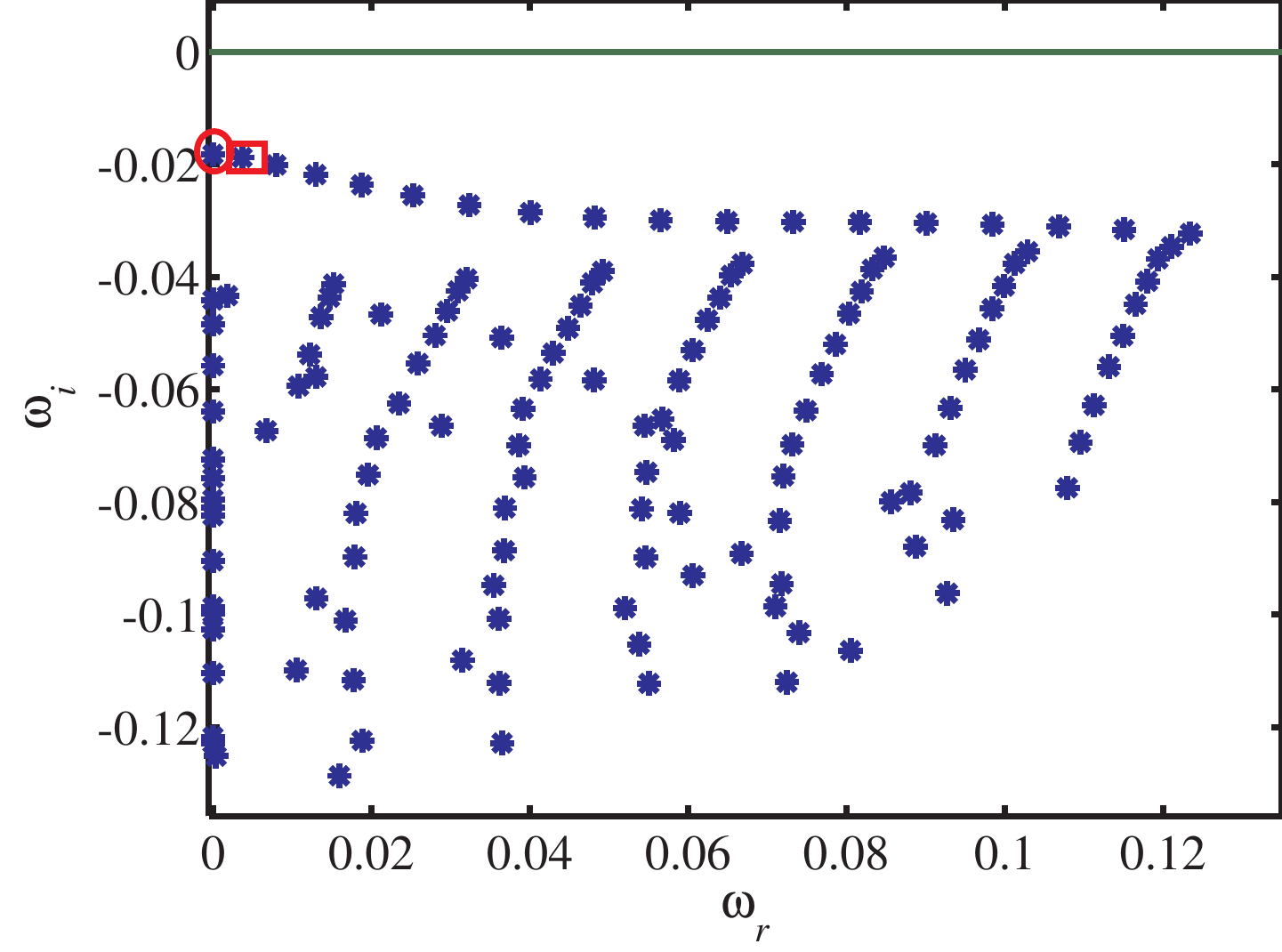}}          
\caption{Eigenspectrum for azimuthal wavenumber N=0 (axisymmetric mode) and Re=383.}
\label{sp_0_383}
\end{figure}
\begin{figure}  
\centerline {\includegraphics [height=1.0in, width=4.5in, angle=0]
                              {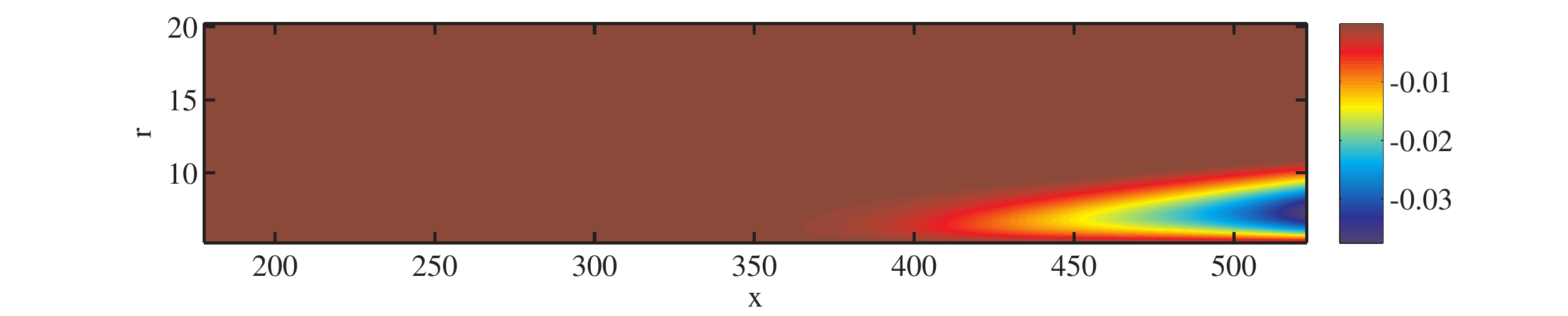}}  
\centerline {\includegraphics [height=1.0in, width=4.5in, angle=0]
                              {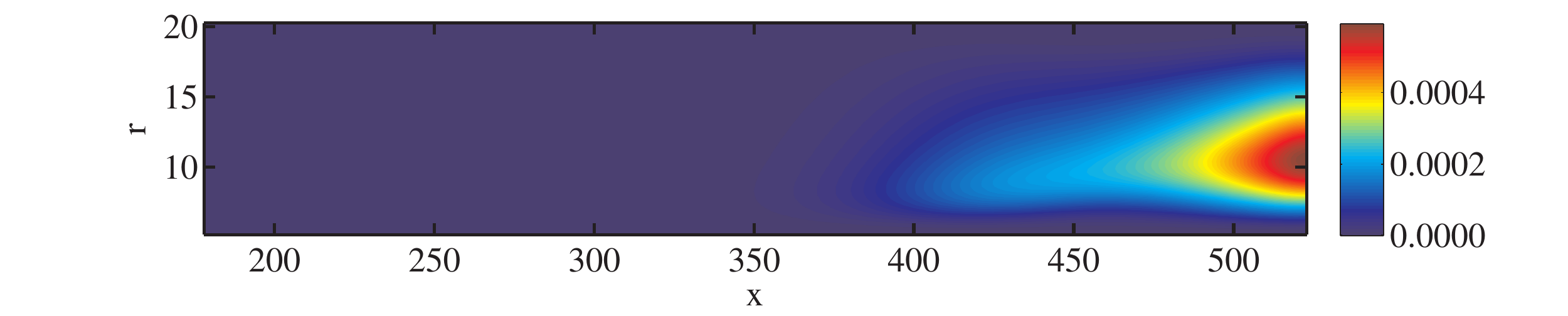}}
\caption{Contour plot of stationary mode (a) streamwise velocity and 
        (b) normal velocity for the eigenvalue $\omega=0.0-0.01818i$, 
        as marked by circle in figure \ref{sp_0_383} .} 
\label{cont_0_383_s}
\end{figure}
\begin{figure} 
\begin{center}
\includegraphics [height=1.20in, width=2.0in, angle=0]
                 {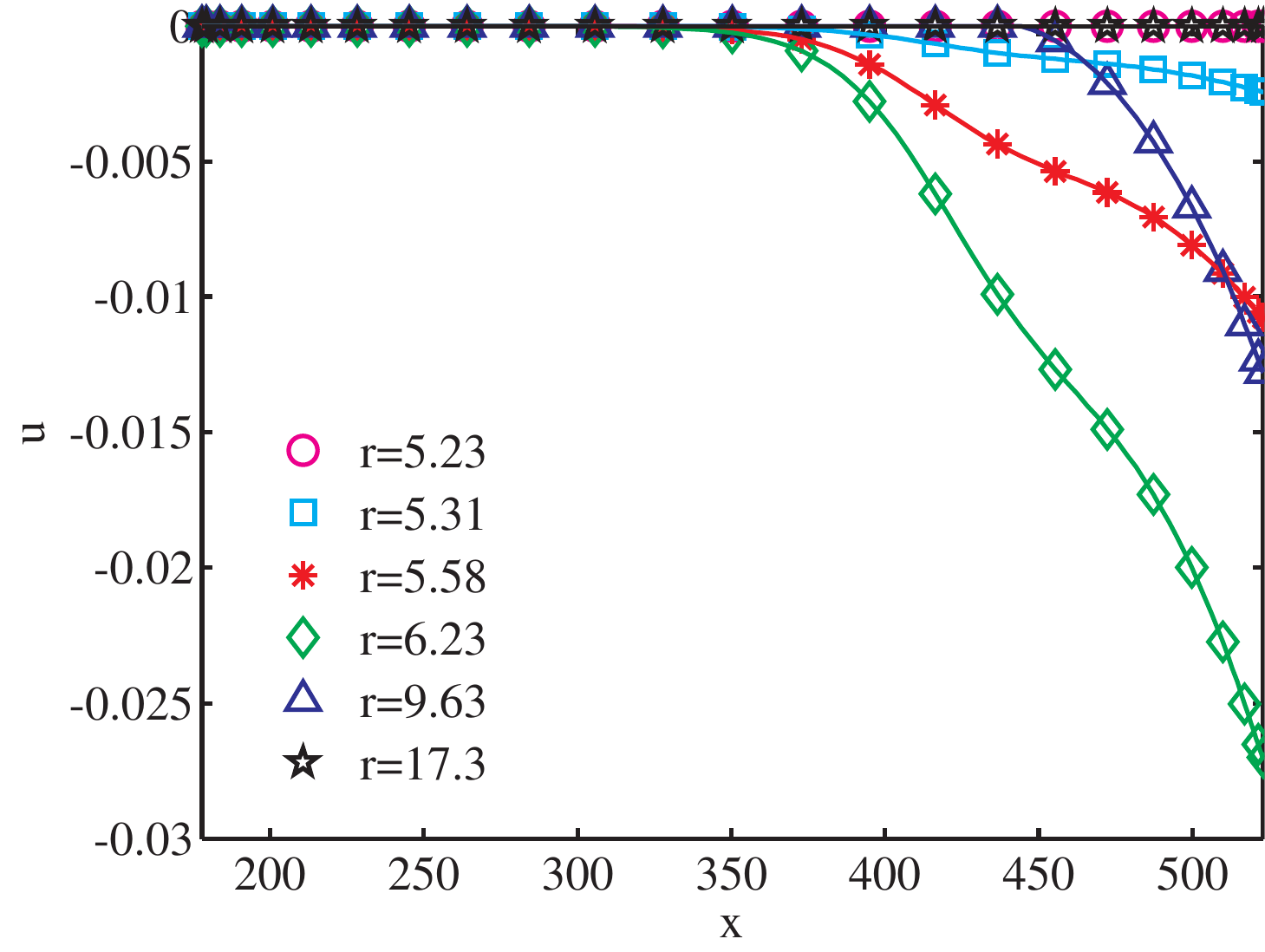} 
\hspace{2px}  
\includegraphics [height=1.25in, width=2.0in, angle=0]
                 {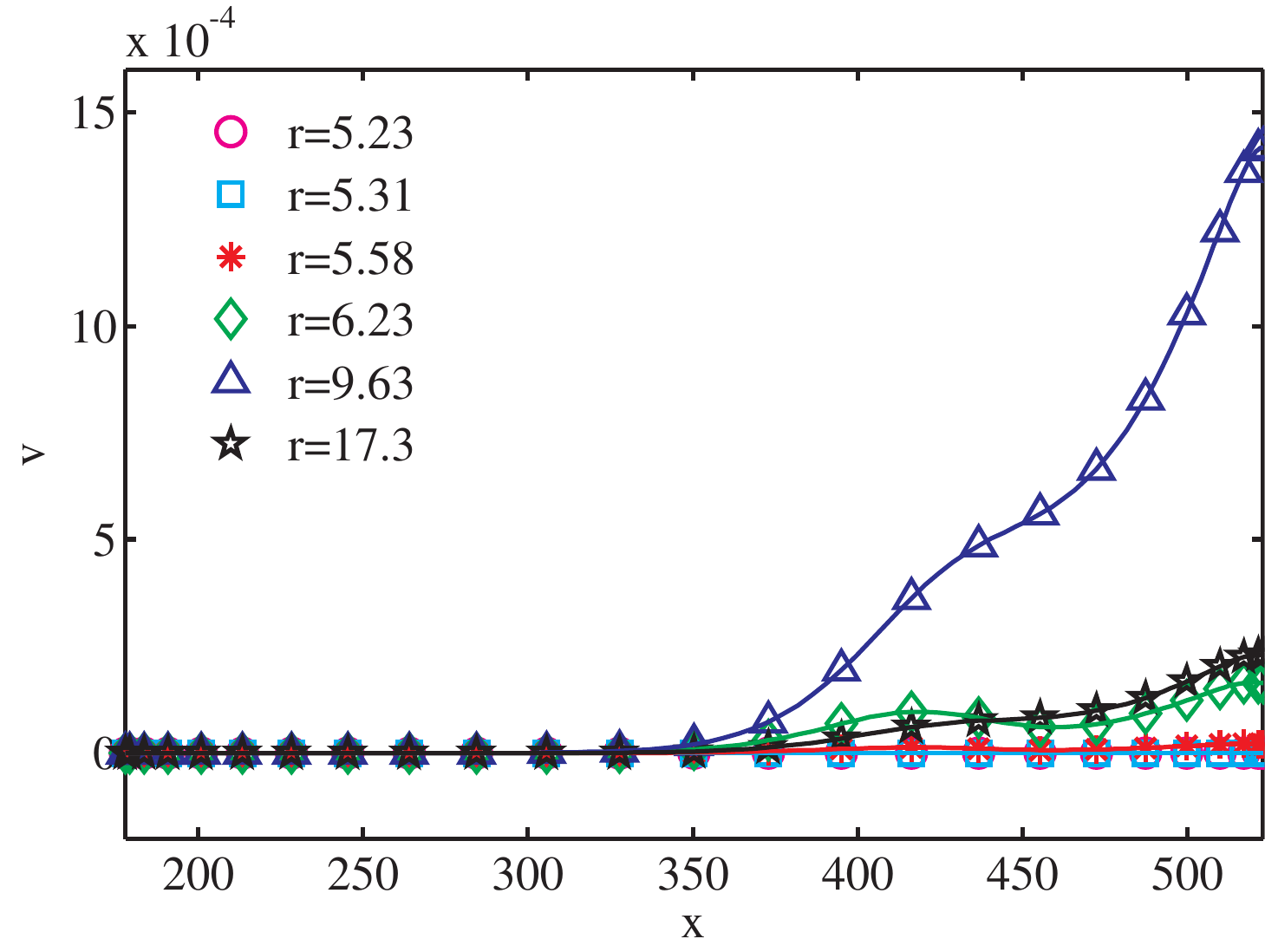}   
\caption {Variation of disturbance amplitudes in streamwise direction 
         for stationary mode (a) streamwise velocity (u) and 
         (b) normal velocity(v) for the eigenvalue $\omega=0.0-0.01818i$, 
         as marked by circle in figure \ref{sp_0_383} .} 
\label{plot_0_383_s}
\end{center}
\end{figure}
 \subsection {Axisymmetric mode}
 Figure \ref {sp_0_383} shows the eigenspectrum for the axisymmetric mode (N=0)
 with Reynolds number, $Re=383$ .
 In figure \ref{sp_0_383}, the eigenmodes marked by circle and square
 are  stationary ($\omega_r=0$) and oscillatory($\omega_r>0$) 
 modes respectively.  
 The stationary mode has a complex frequency $\omega=0-0.01818i$. 
 This global mode is temporally  stable because $\omega_i < 0$ 
 and hence disturbances decay in time.
 Figure \ref{cont_0_383_s} shows the two dimensional spatial structure 
 of the streamwise(u) and wall normal (v) disturbance amplitudes for stationary 
 mode $\omega=0-0.01818i$.  
 The magnitudes of the velocity disturbances is zero at the inlet,
 as it is the inlet boundary condition.
 As the fluid particles move towards the downstream, the disturbances evolve 
 monotonically in time within the domain and progresses towards the downstream. 
 The magnitude of the u disturbance amplitudes is one order higher than that of 
 v disturbance amplitudes.
 The amplitude structure has opposite sign for u and v.
 The spatial structure of the amplitudes grow in size and magnitude towards the downstream.
 Figure \ref {plot_0_383_s} shows the variation of u and v disturbance amplitudes 
 in streamwise direction at different radial locations.
 The magnitude of the disturbance amplitudes are very small near the cylinder
 surface due to viscous effect, increases in radial direction and vanishes at far-field.\\
 \begin{figure}      
  \centerline {\includegraphics [height=1.0in, width=4.5in, angle=0]
              {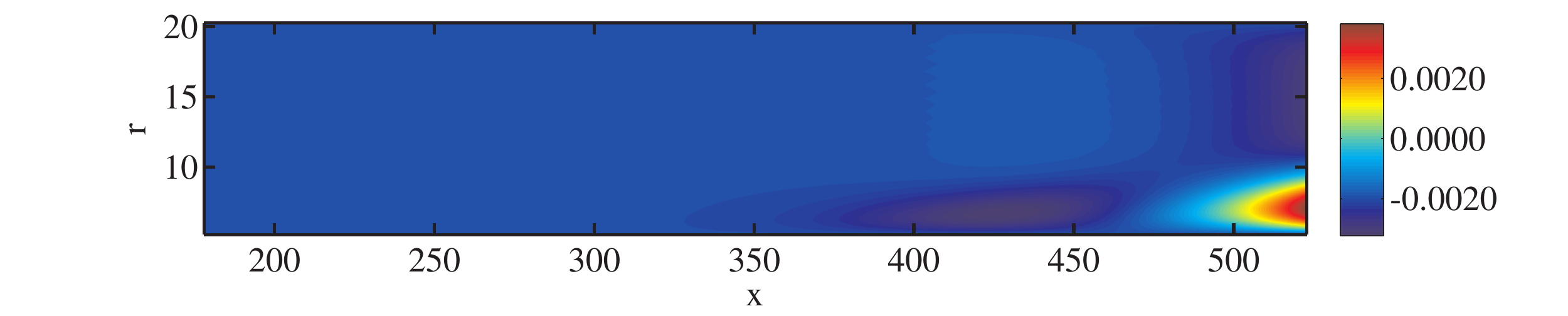}}   
  \centerline {\includegraphics [height=1.0in, width=4.5in, angle=0]
              {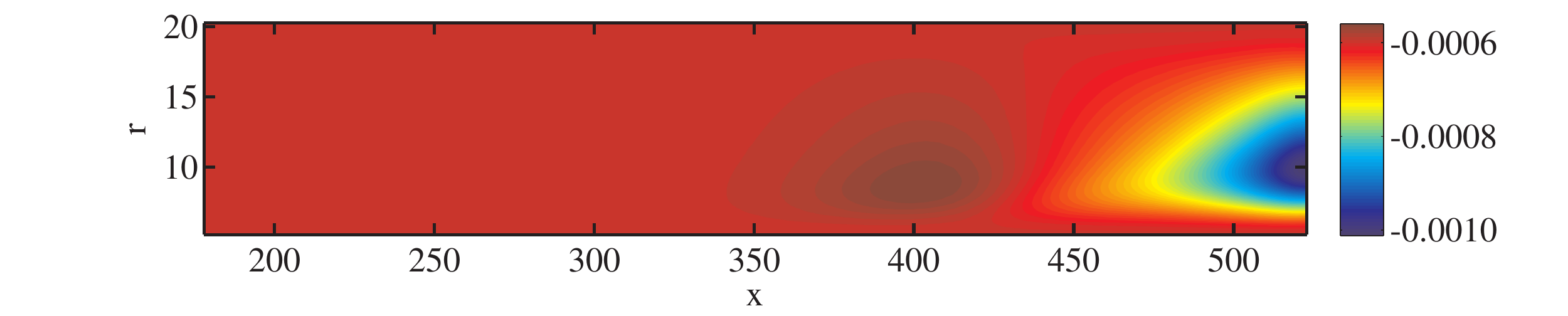}}
  \caption{Contour plot of oscillatory mode (a) streamwise velocity and 
          (b) normal velocity for the eigenvalue $\omega=0.003689-0.01874i$, 
          as marked by square in figure \ref{sp_0_383} .} 
  \label{cont_0_383_o}
  \end{figure}
  \begin{figure} 
  \begin{center} 
  \includegraphics [height=1.25in, width=2.1in, angle=0]
                   {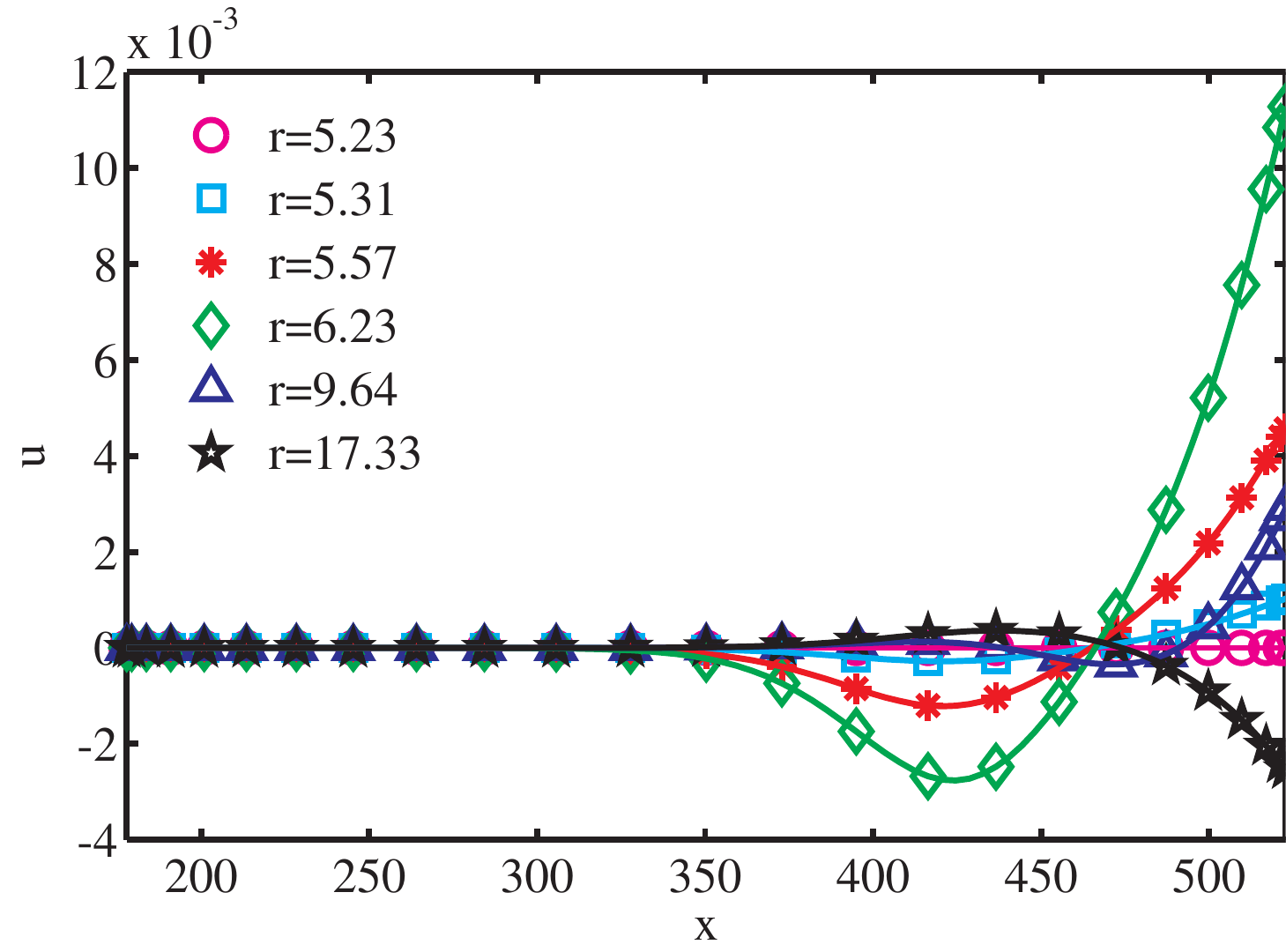}   
  \hspace{1px}
  \includegraphics [height=1.25in, width=2.1in, angle=0]
                   {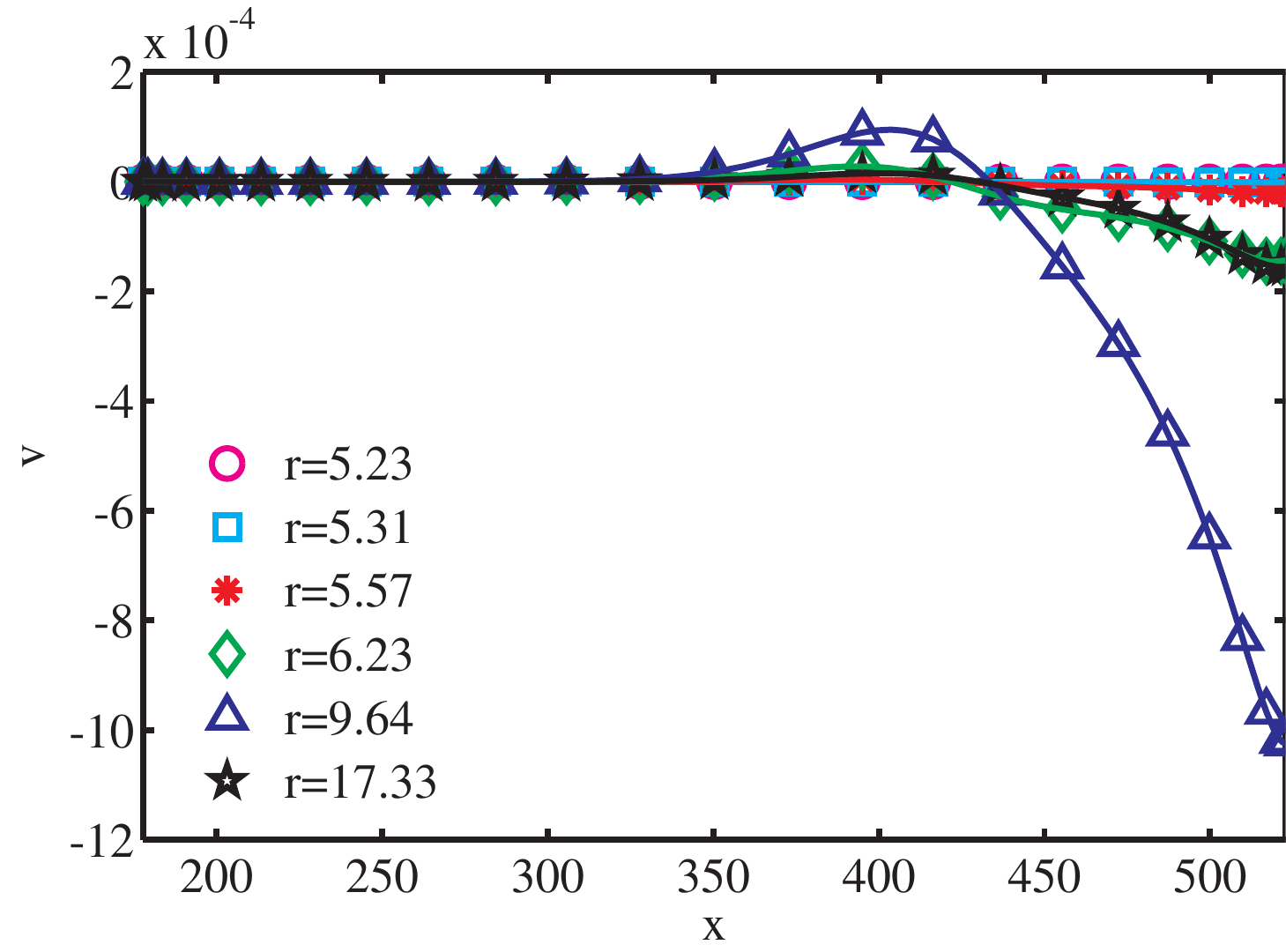}   
  \caption {Variation of disturbance amplitudes in streamwise direction 
           for oscillatory mode (a)streamwise velocity (u) and 
           (b) normal velocity(v) for the eigenvalue $\omega=0.003689-0.01874i$, 
           as marked by square in figure \ref{sp_0_383} .} 
  \label{plot_0_383_o}
  \end{center}
  \end{figure}
 The most unstable oscillatory mode has an eigenvalue $\omega=0.003689-0.01874i$. 
 The flow is stable for this global mode because $\omega_i < 0$. 
 The spatial structure of the above global eigenmode for u and v disturbance amplitudes  
 are not monotonic in nature.The distribution in spatial directions is show in 
 figure \ref{cont_0_383_o} . 
 The amplitudes of the velocity disturbances is zero at the inlet  as it is the imposed 
 boundary condition at the inlet .
 However, in the streamwise direction the disturbances grow in magnitudes  and 
 it contaminates the flow field towards the downstream.
 The magnitudes of the disturbance amplitudes grows exponentially as they move 
 further towards the downstream  and hence flow is convectively unstable \cite{Rodriguez}.
 However, its variation in the normal direction is different from that of a 
 stationary mode, as seen in the respective figures. 
 The magnitude of the normal disturbance amplitudes is one order less than 
 that of the streamwise component.\\
 Figure \ref{plot_0_383_o} presents variation of disturbance amplitudes in streamwise 
 direction at various radial locations. 
 It shows that the most of the disturbances grow in magnitude as they move towards downstream. 
 The magnitude of the disturbance amplitudes  are very small near the wall due to viscous effect,
 gradually increases in radial direction and finally vanishes at the far-field. 
 It shows that disturbances evolve within the flow-field in time and grow in magnitude and size 
 while moving towards the downstream.\\
 \subsection{Effect of frequency on eigen function}
\begin{figure}      
  \centerline{\includegraphics [height=1.0in, width=4.5in, angle=0]
                               {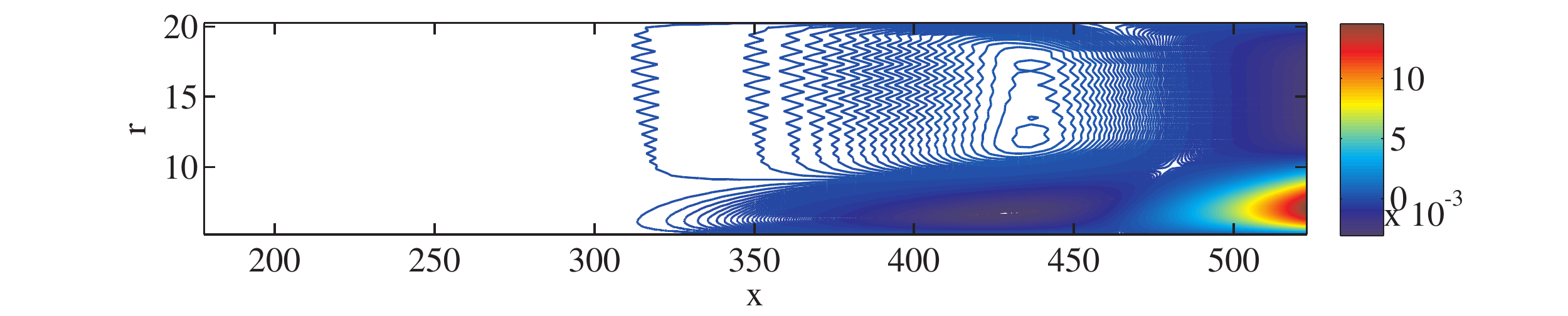}}   
  \centerline{\includegraphics [height=1.0in, width=4.5in, angle=0]
                               {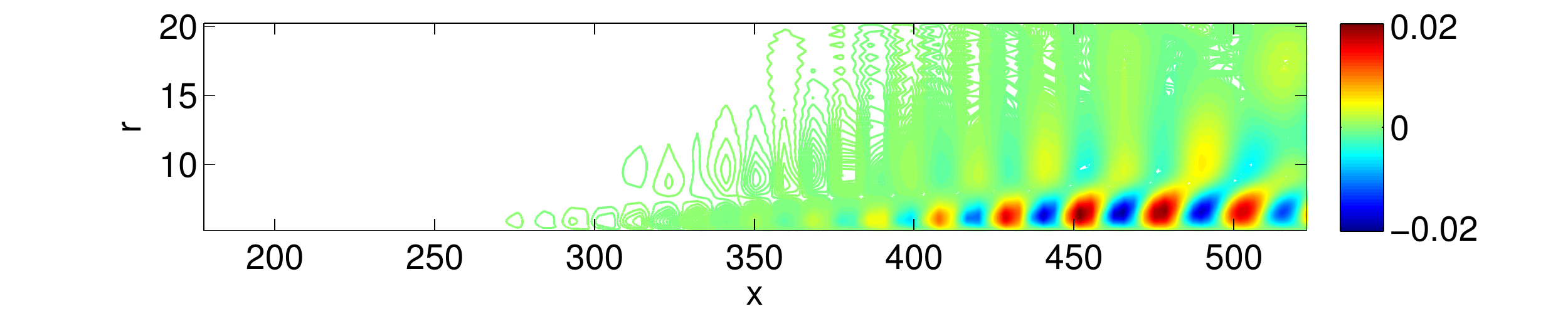}}
  \caption{Contour plot of the real part of u for the two different eigenmodes 
          associated with the frequencies (a) $\omega_r$=0.003689 (b)$\omega_r$=0.1234.  
          The corresponding Reynolds number and azimuthal wave-number are Re=383 
          and N=0 respectively.}
  \label{cont_0_383_omegar}
  \end{figure}

  Figure \ref{cont_0_383_omegar} shows  contour plot of the real part of 
  streamwise disturbance velocity of two different eigen modes with frequency 
  $\omega_r$=0.003689 \& $\omega_r$=0.1234 for N=0 and Re=383. 
  The streamwise domain length is 345.
  The disturbances is seen to evolve in the vicinity of the wall and increasing the amplitudes 
  when moving  towards downstream. The typical length scale of the the wavelet 
  structure decreases with the increases in frequency($\omega_r$) .\\
  \subsection {Effect of transverse curvature}
  The body radius of the cylinder is another important length scale
  in case of the axisymmetric boundary layer in addition to  the boundary 
  layer thickness ($\delta^*$).
  The inverse of the body radius is called the transverse surface curvature(S).
  The flat plate boundary layer is a special case of a zero transverse curvature. 
  It is normalised with the displacement thickness ($\delta^*$) at the inlet.
  \begin{figure}  
  \begin{center}    
  \includegraphics [height=1.5in, width=2.25in, angle=0]{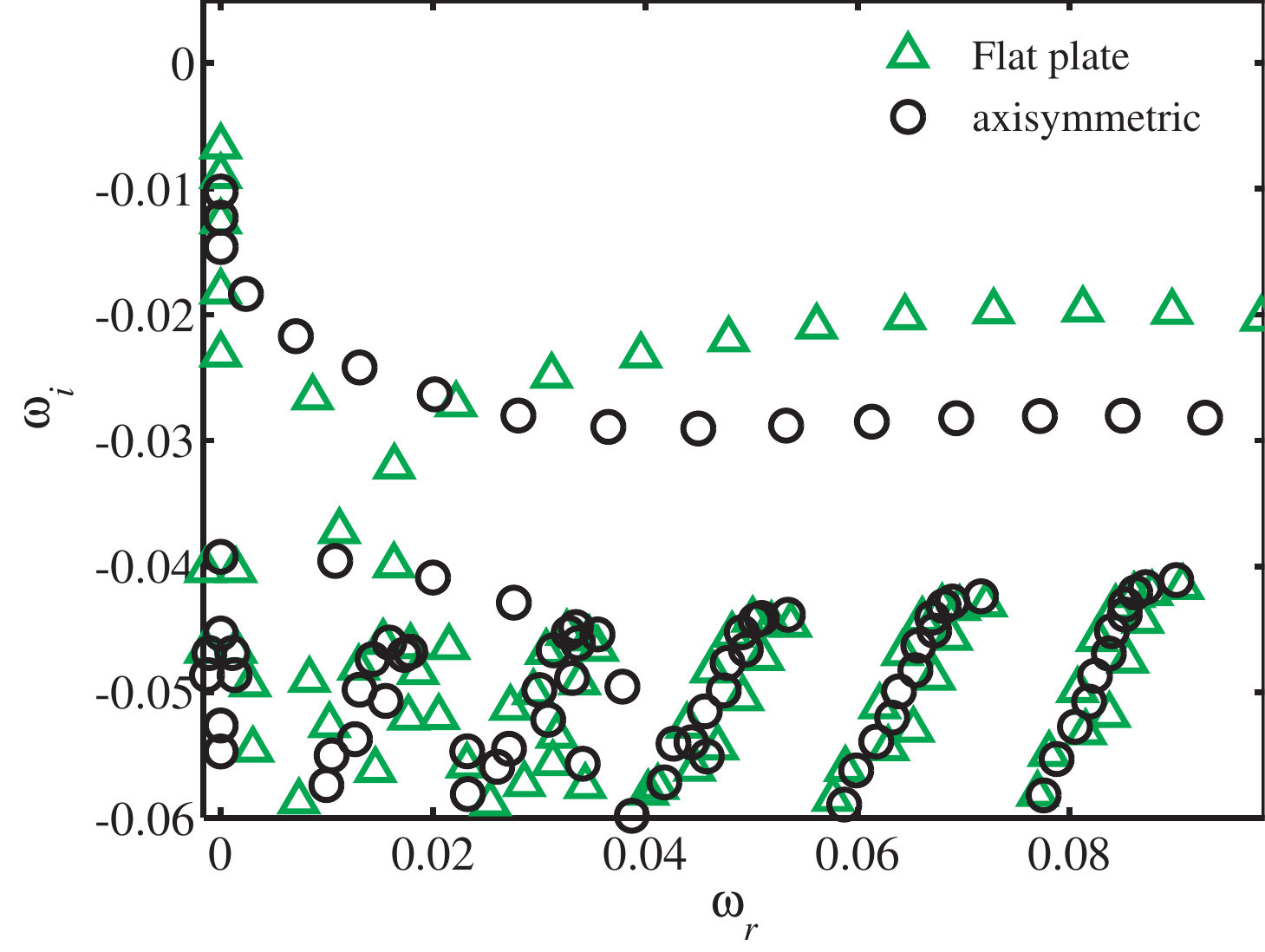}   
  \includegraphics [height=1.5in, width=2.25in, angle=0]{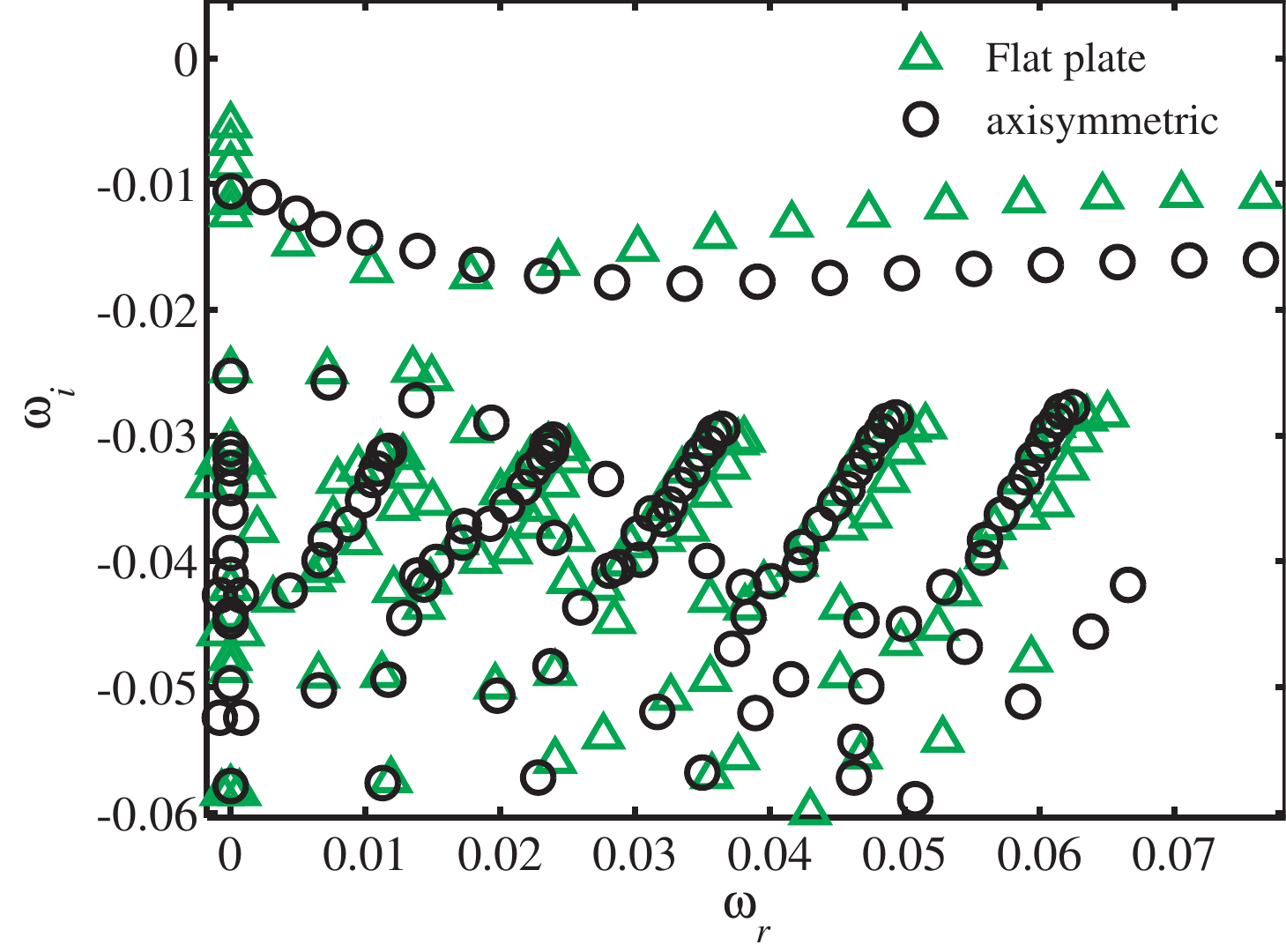}
  \end {center}
  \begin{center}
  \includegraphics [height=1.5in, width=2.25in, angle=0]{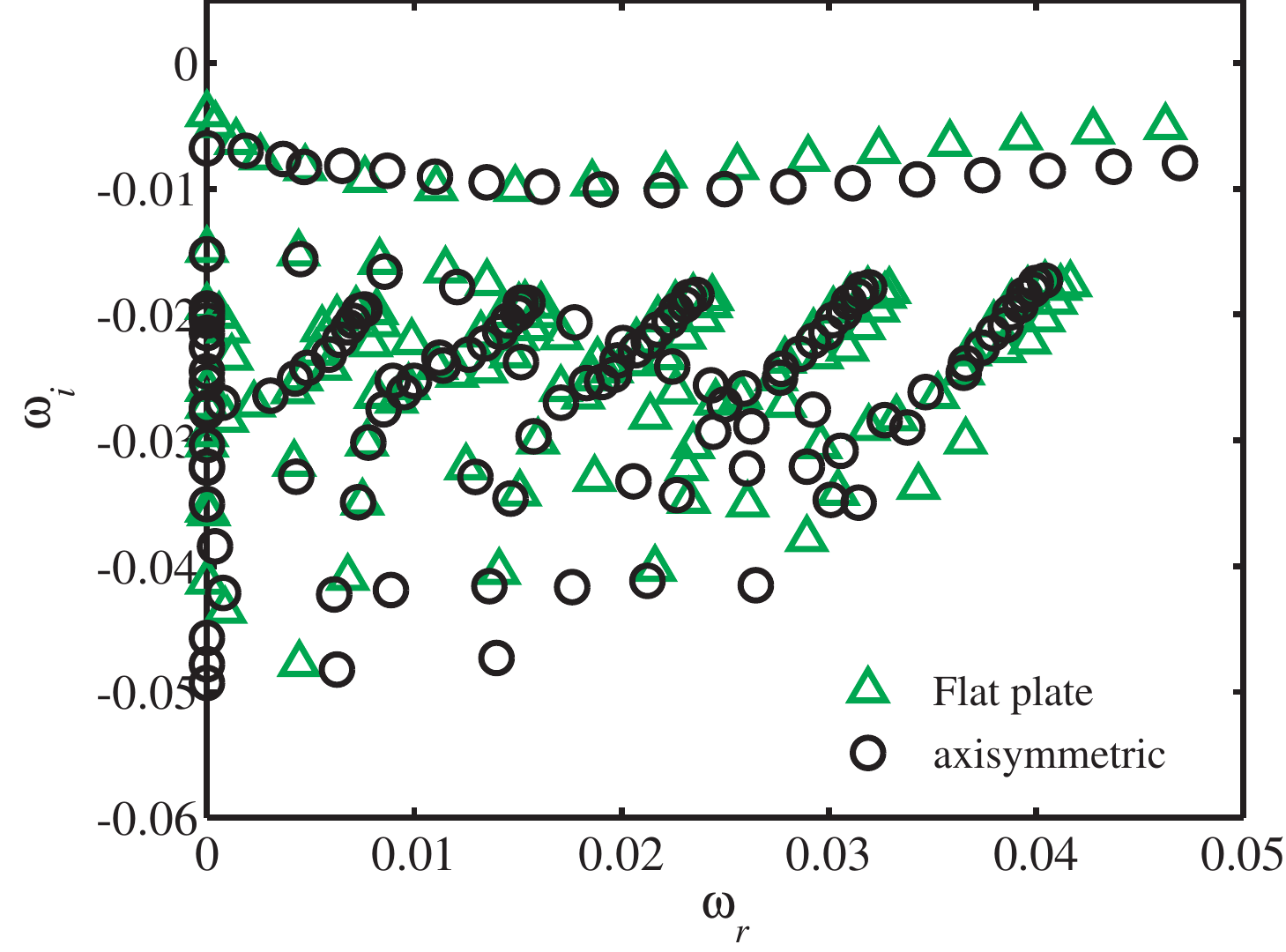} 
  \caption {Comparison of eigenspectrum for axisymmetric (N=0) and 2D flat plate 
           ($\beta=0$) boundary layer for $L_{x=345}$ for three different Reynolds
            number (a) Re=383 (b) Re=557 and (c) Re=909.} 
  \label{sp_plate_cyl}
  \end{center}
  \end{figure}
  In case of the axisymmetric boundary layer it has significant effect on the characteristic
  of base flow as well as disturbance flow quantities.
  In order to understand the transverse curvature effect on the stability characteristic
  of the axisymmetric boundary layer,the stability characteristics are compared  with 
  the flat plate boundary layer(zero transverse curvature) at the same Reynolds number 
  and domain size.
  The Reynolds number is computed based on the displacement thickness and free-stream 
  velocity at the inlet.
  The eigen spectrum, spatial amplification rate, effect of the domain length on 
  the distribution of the spectrum and eigen functions are compared at the same 
  Reynolds number. 
  \begin{figure}  
  \begin{center}    
  \includegraphics [height=1.25in, width=1.75in, angle=0]{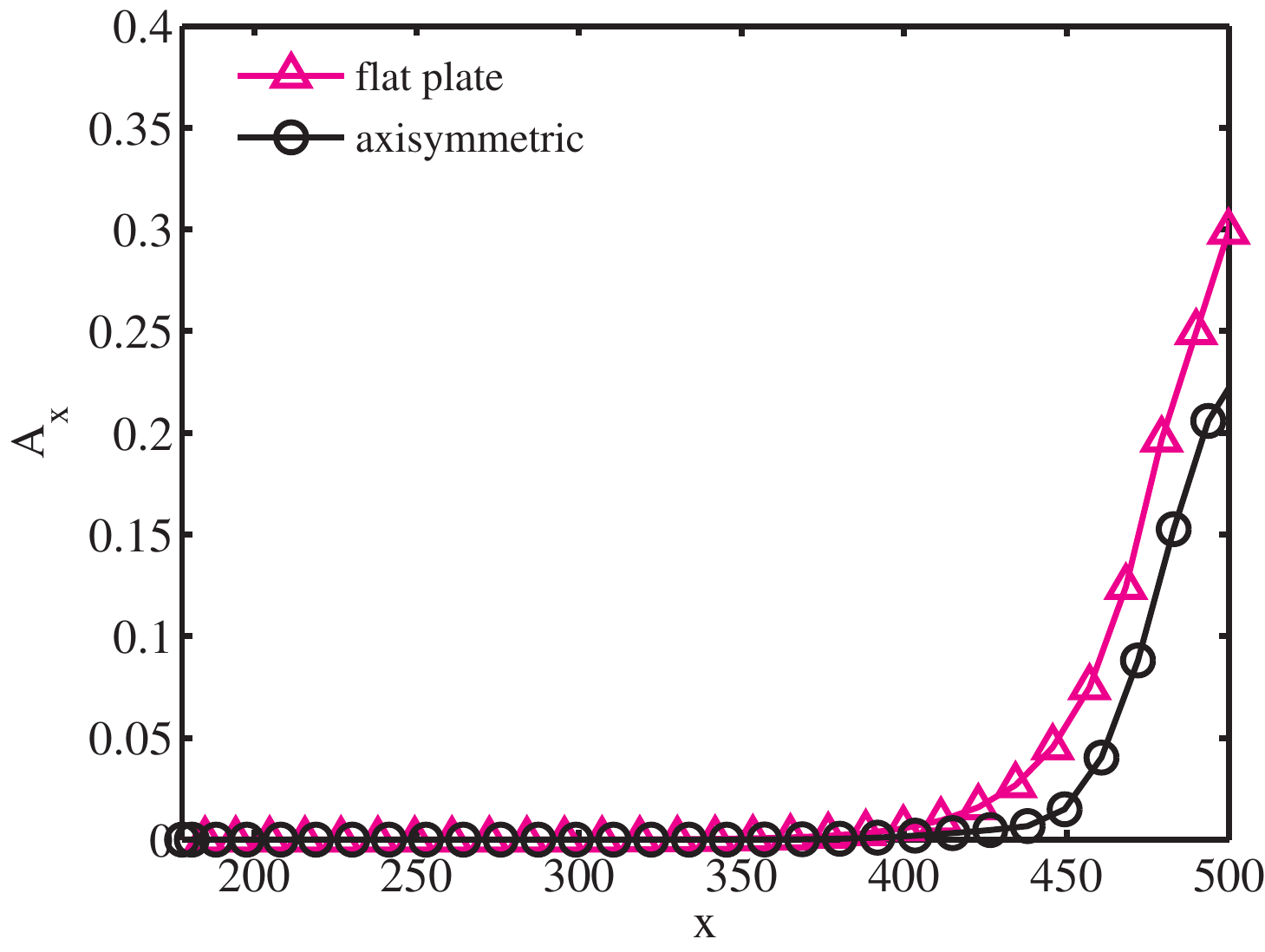}   
  \includegraphics [height=1.25in, width=1.75in, angle=0]{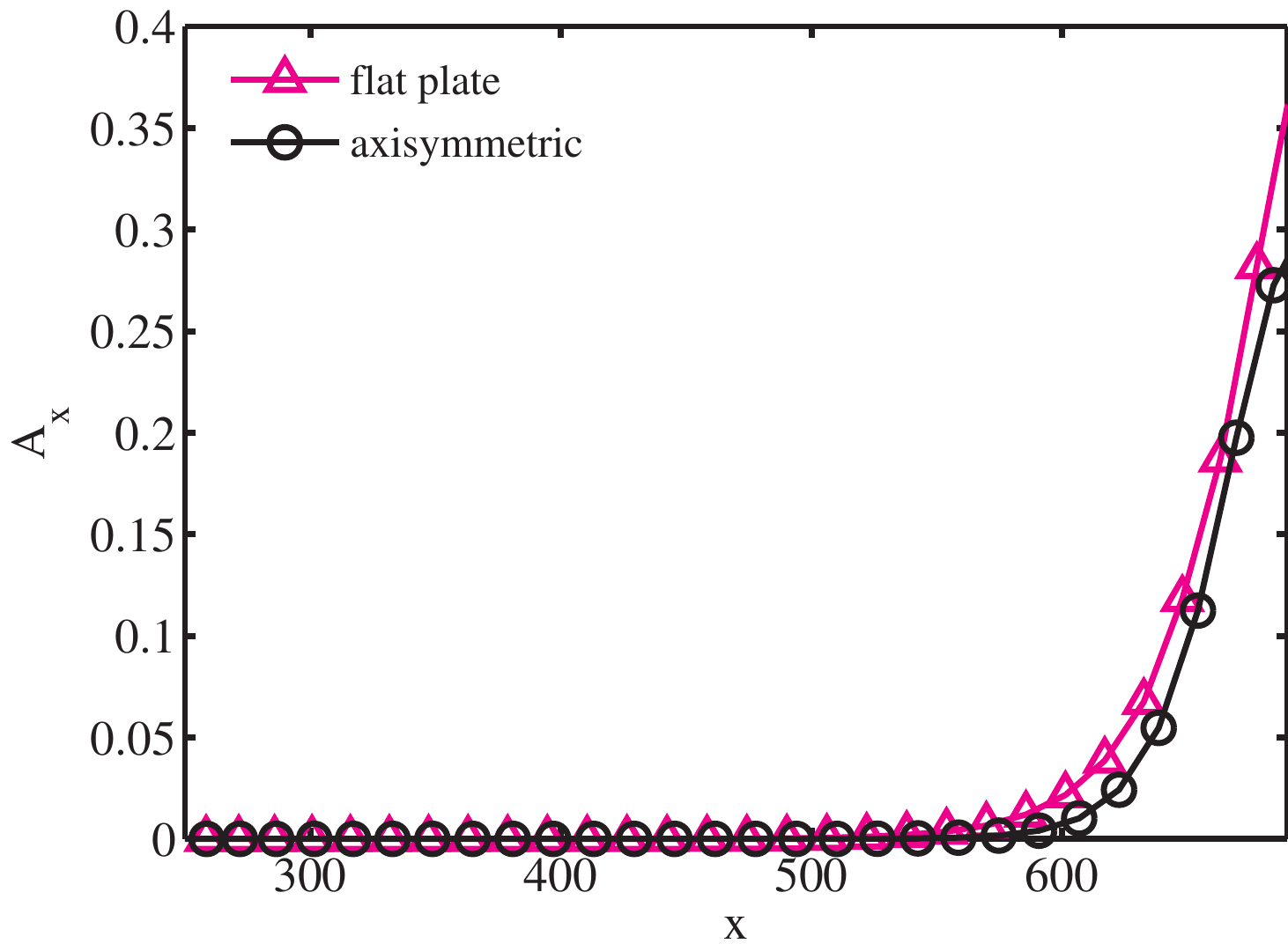}
  \end{center}
  \begin{center}
  \includegraphics [height=1.25in, width=1.75in, angle=0]{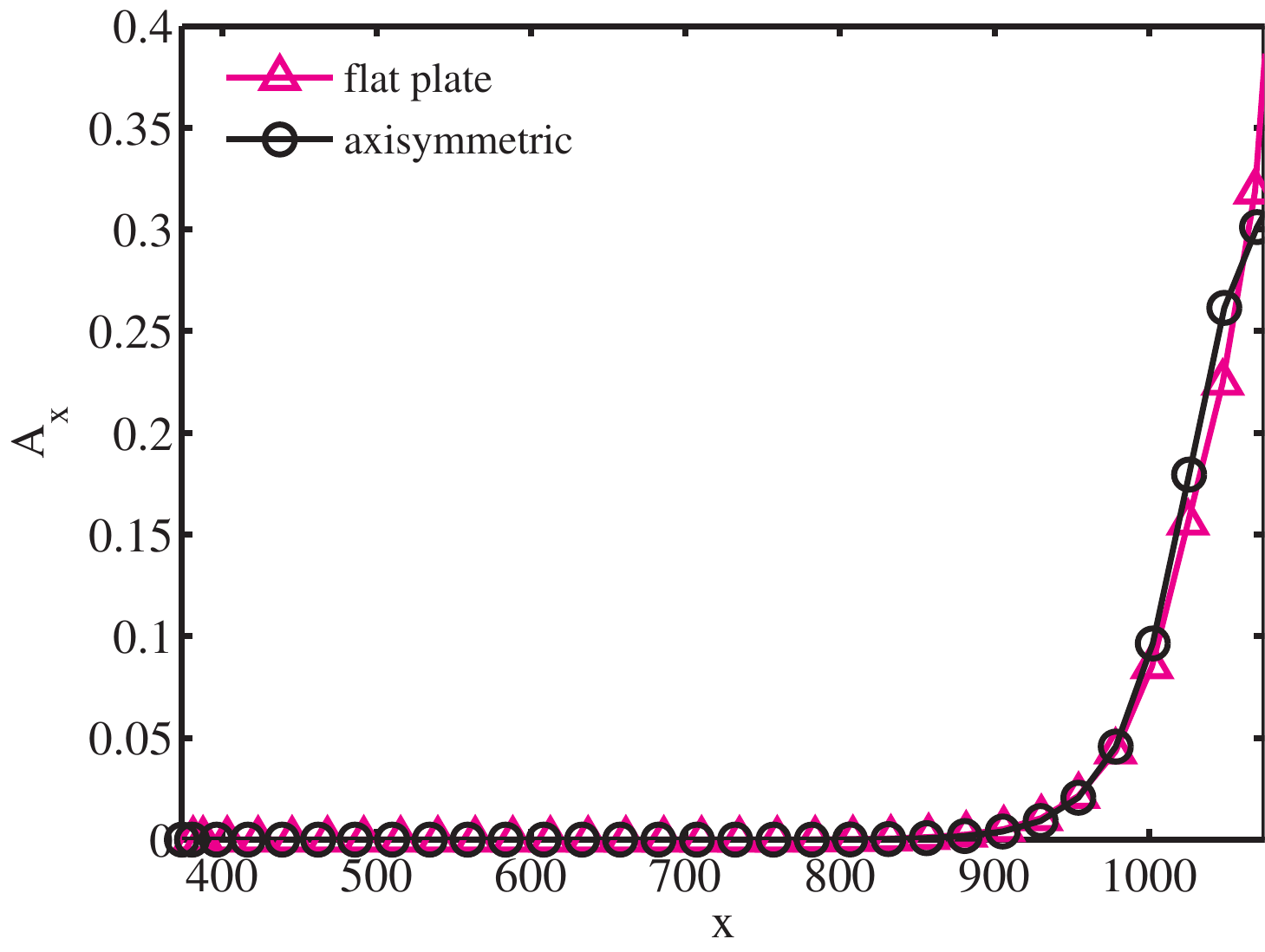}      
  \caption  {Comparison of spatial amplification rate of axisymmetric (N=0) 
            and 2D flat plate ($\beta=0$)boundary layer for three different 
            Reynolds number (a) Re=383 (b) Re=557 and (c) Re=909.} 
  \label{Ax_plate_cyl}
  \end{center}
  \end{figure}
  Figure \ref{sp_plate_cyl} shows the comparison of eigen spectrum for three
  different Reynolds numbers 383, 557 and 909 (based on the body radius of the 
  cylinder these are $2000$, $4000$ and $10000$) with  $L_{x}$=345 for axisymmetric
  and flat plate boundary layer. 
  The comparison is limited for the discrete part of the spectrum only.  
  It has been observed that at small Reynolds number the damping rates of eigen
  modes are higher ($\omega_{i}$ is smaller) for axisymmetric boundary layer 
  than that of flat plate boundary layer.
  The difference of damping rate in eigen modes for axisymmetric and flat plate 
  boundary layer reduces with the increase in Reynolds number.
  The eigen modes of axisymmetric boundary layer approaches to 
  eigen modes of flat plate boundary layer at higher Reynolds number.
  This is primarily due to the effect of transverse curvature only.
  The effect of transverse curvature is significant at low Reynolds number.
  The effect of transverse curvature reduces with the increase in Reynolds number.
  Thus, it proves that the global modes are more stable at low Reynolds number 
  for axisymmetric boundary layer.\\
  \begin{figure}
  \begin{center} 
  \includegraphics[height=1.25in, width=1.75in, angle=0] {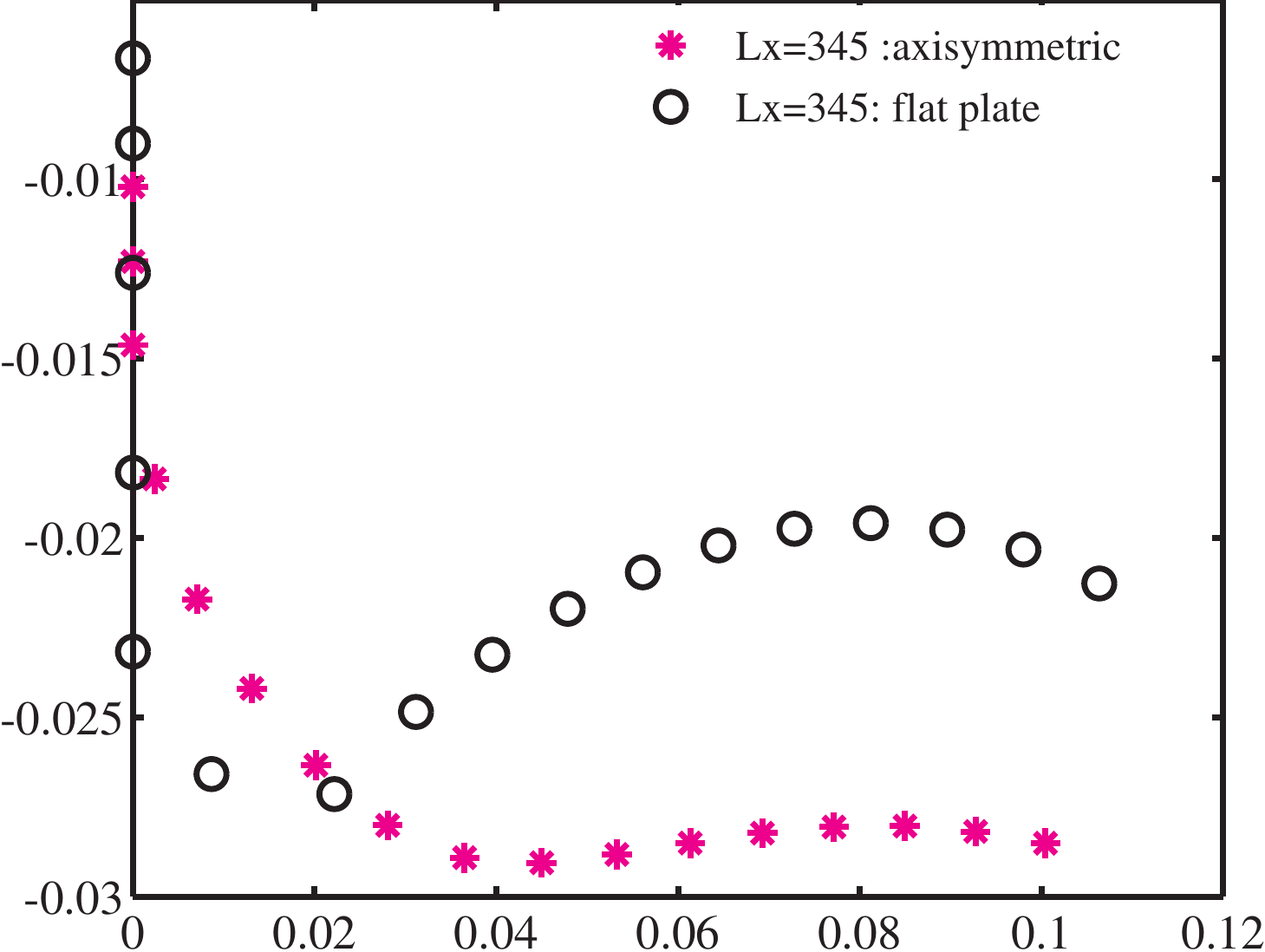}          
  \includegraphics[height=1.25in, width=1.75in, angle=0] {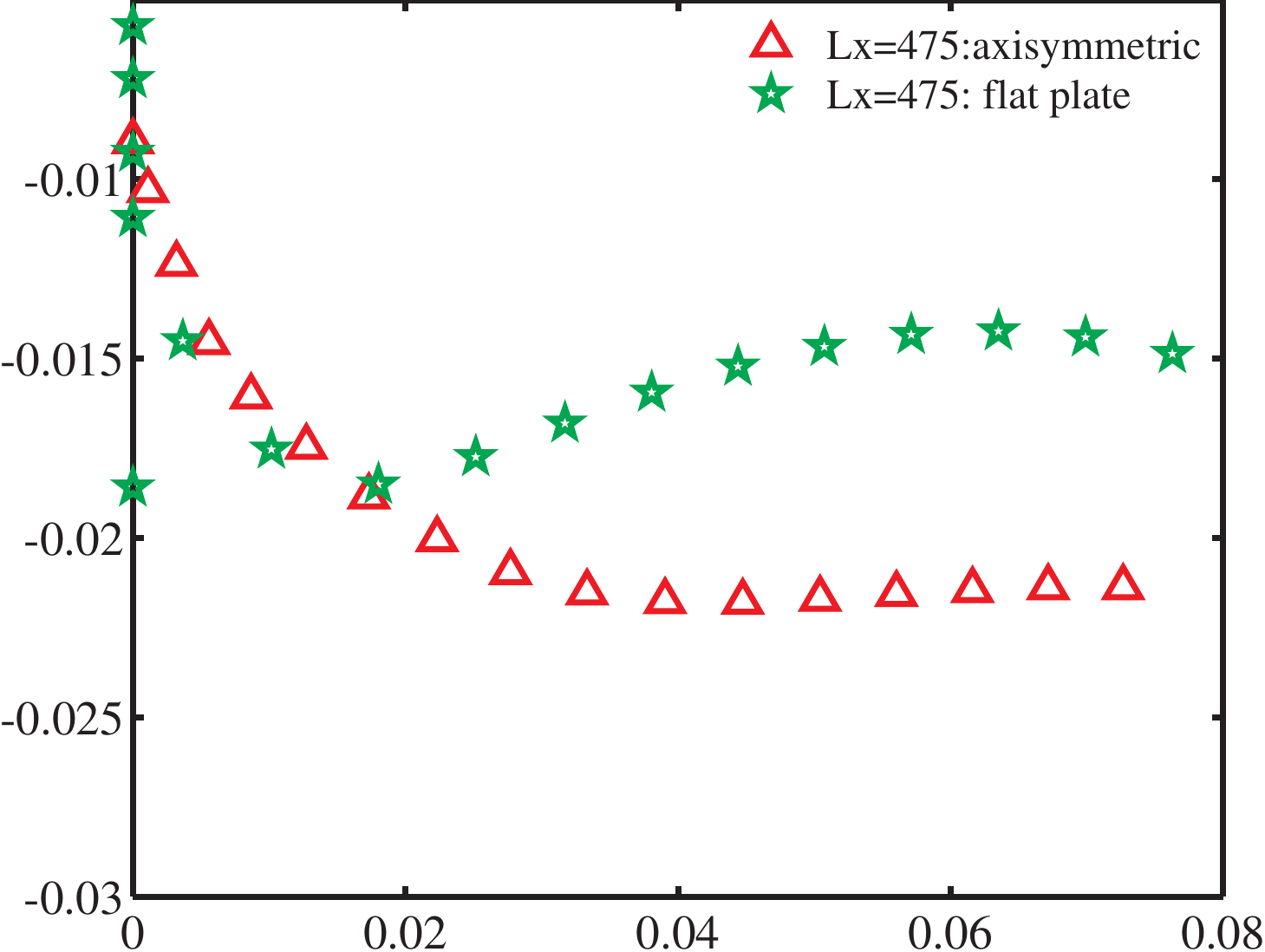} 
  \end{center}
  \begin{center}  
  \includegraphics[height=1.25in, width=1.75in, angle=0] {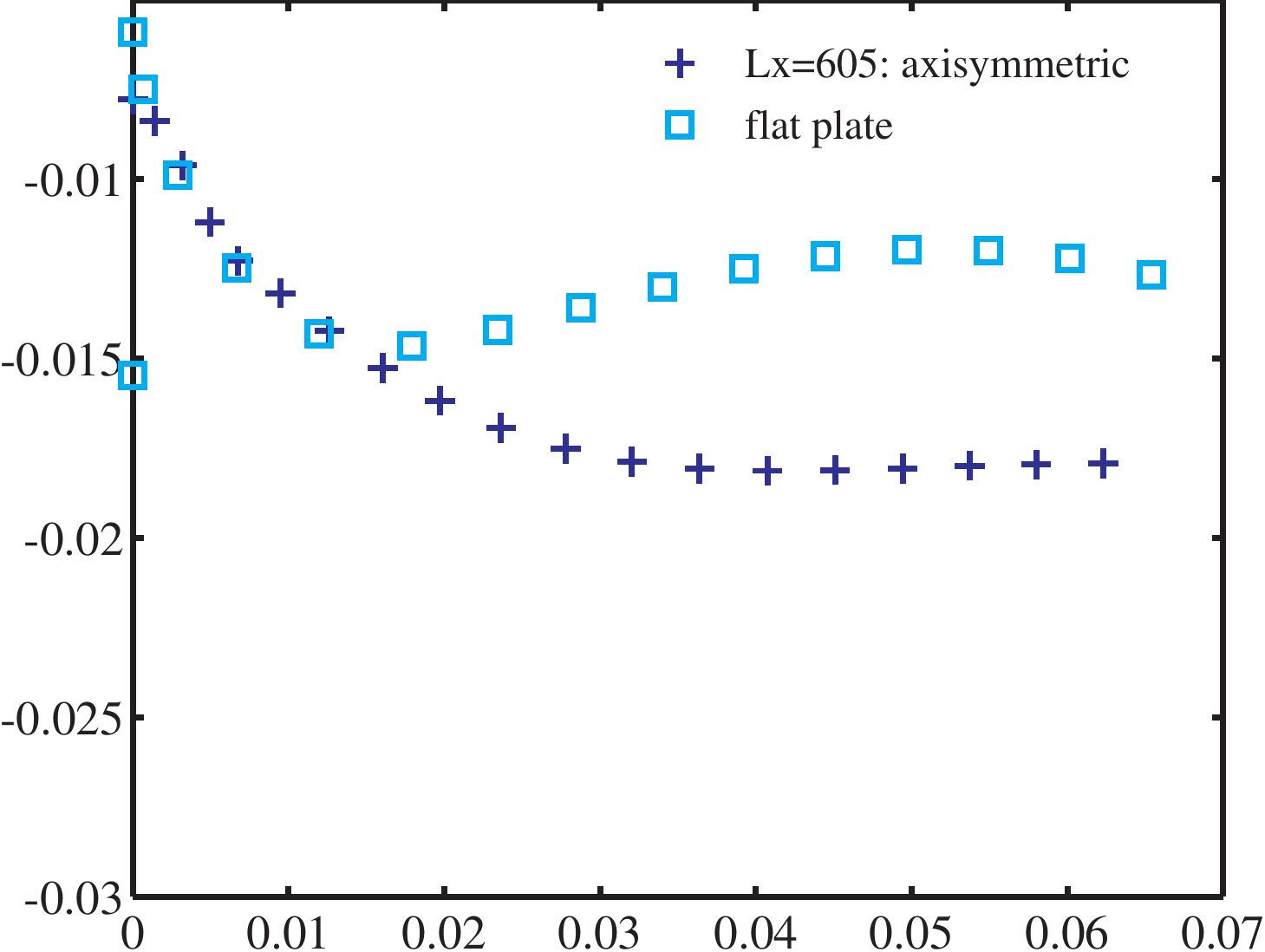}  
  \caption{(a)Comparison of eigen spectrum for different stream-wise domain 
             length for Re=383 for axisymmetric (N=0) and 2D flat plate 
             boundary layer($\beta=0)$.}
  \label{sp_Lx_383} 
  \end{center} 
  \end{figure}
  Figure \ref{Ax_plate_cyl} shows the comparison of spatial amplification rates
  ($A_{x}$)at three different Reynolds number for both the boundary layers. 
  The least stable mode is selected to compute the spatial amplification rate.
  The comparison shows that at low Reynolds number, ($A_{x}$)is higher for flat plate 
  boundary layer.
  The ($A_{x}$) for axisymmetric boundary layer also increases with the increase 
  in Reynolds number and approaches to flat plate boundary layer at higher Reynolds number.\\
 \begin{figure}
 \begin{center} 
 \includegraphics[height=1.5in,width=1.5in,angle=0]{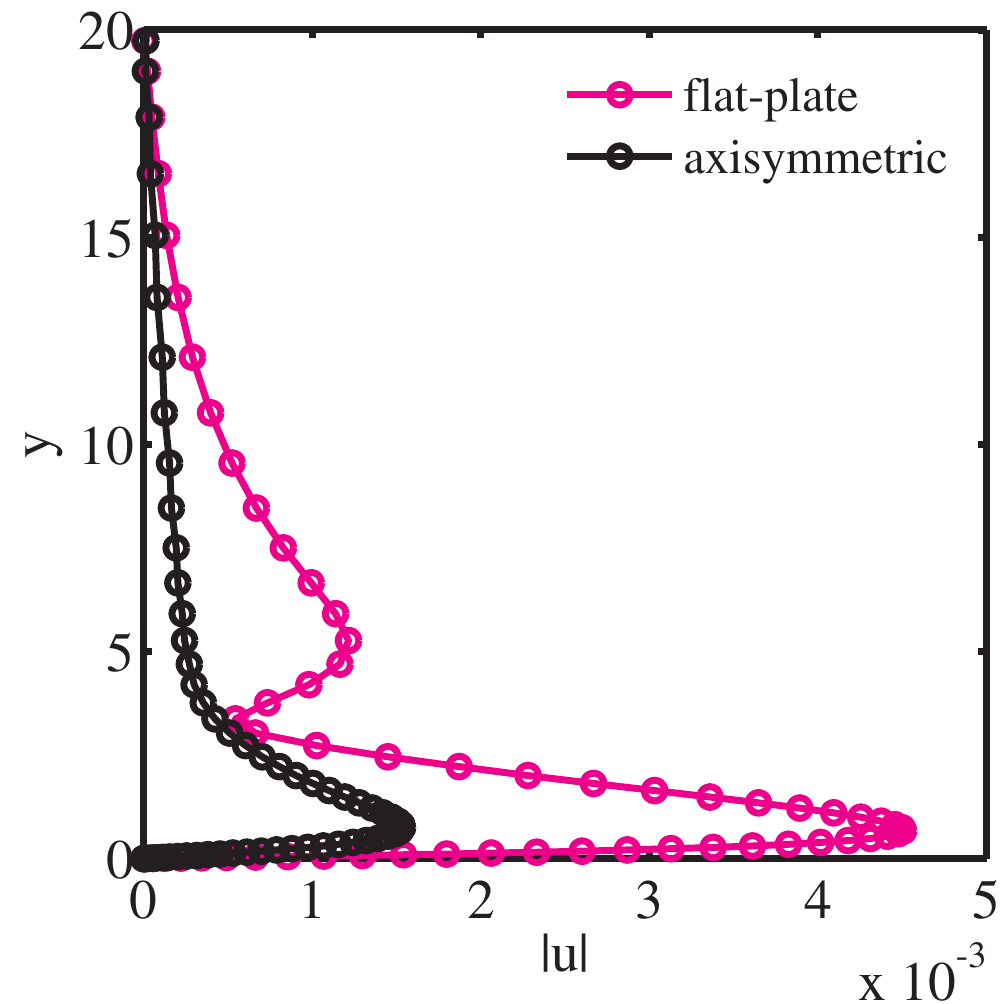}          
 \includegraphics[height=1.5in,width=1.5in,angle=0]{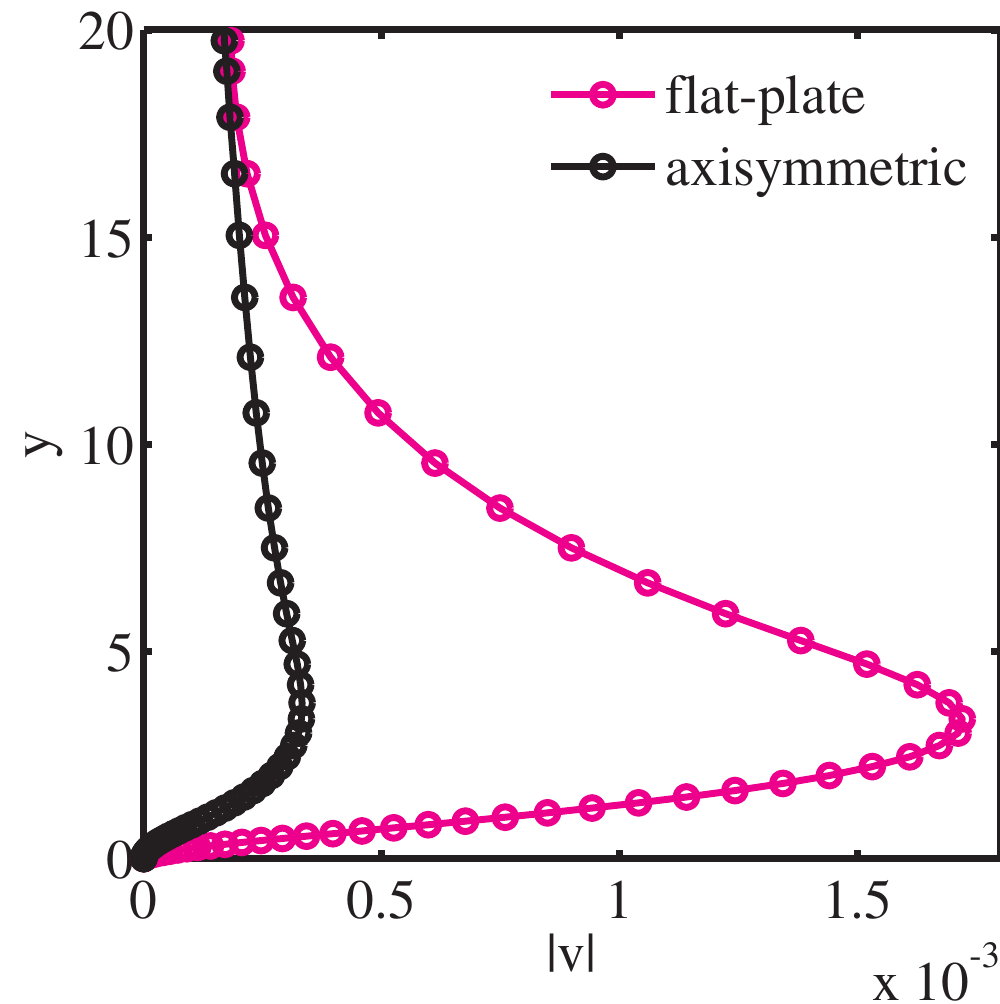}   
  \caption{Comparison of modulus of eigen-function (a) u and (b) v at stream-wise 
           location x=445 for Re=383 for axisymmetric (N=0) and 2D flat plate 
           boundary layer($\beta=0)$.}

 \label{mod_u} 
 \end {center}
 \end{figure}
 Three different families of eigenspectrum are shown in figure \ref {sp_Lx_383} 
 for axisymmetric and flat plate boundary layer. 
 The discrete part of the spectrum is only shown in the figures for the comparison.
 The difference between the spectrum depends on the domain length.
 The distance between two consecutive  frequency reduces with the increase in  
 domain length. 
 To quantify this discretization of the eigen modes for Re=323, different domain lengths 
 were considered with $L_{x1}=345$,$L_{x2}=475$ and $L_{x3}=605$. 
  Figure \ref{mod_u} shows the comparison of modulus of streamwise and normal 
  disturbance velocity u and v respectively for Re=383 at x=445. The magnitude 
  is modulus of disturbances is higher for flat plate plat boundary layer.
  \subsection {Helical mode (N=1) }
  \begin{figure}
  \centerline{\includegraphics[height=1.5in, width=2.2in, angle=0] 
                              {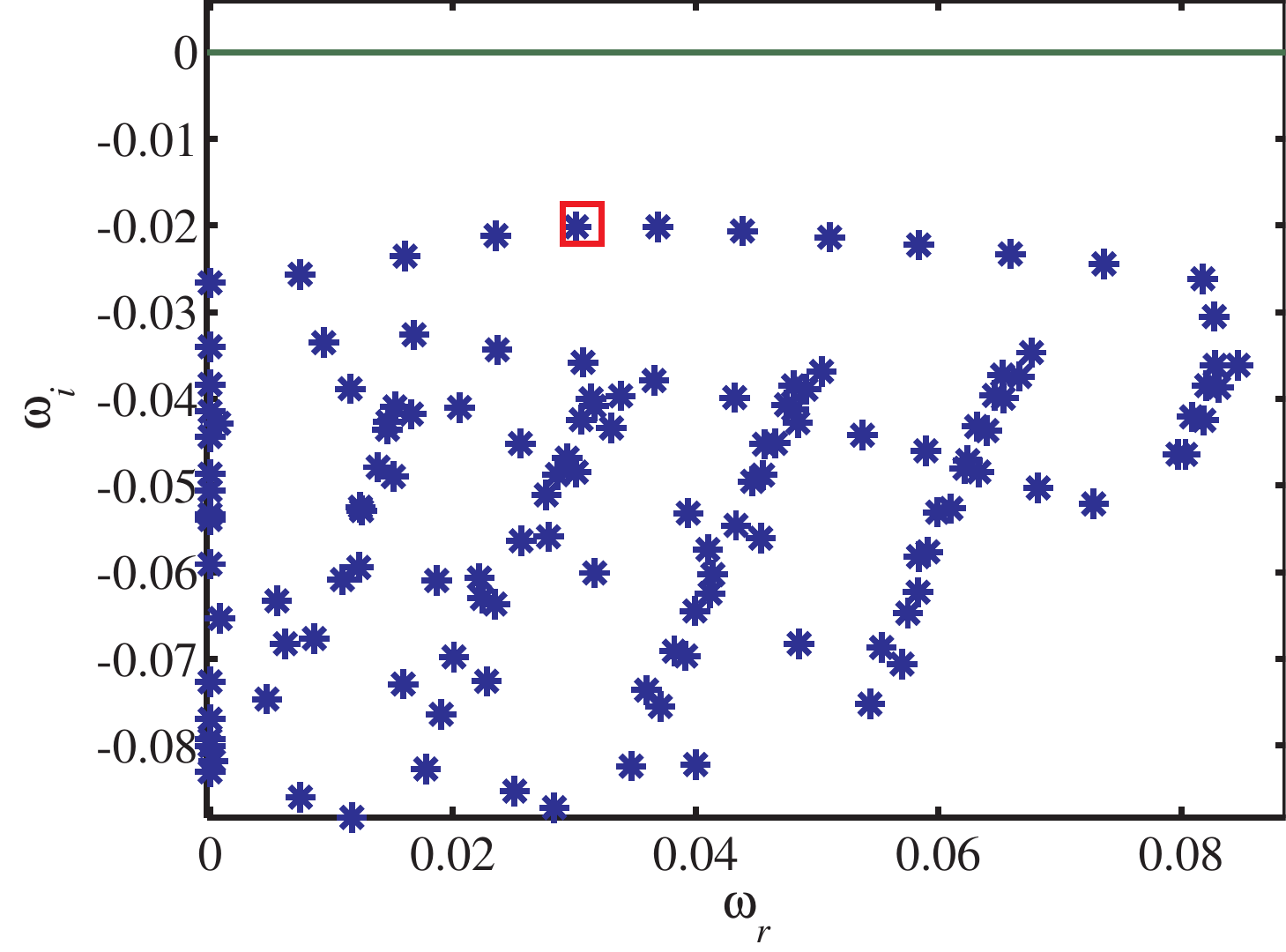}}          
  \caption{Eigenspectrum for azimuthal wavenumber N=1  and Re=383.}
  \label{sp_1_383}
  \end{figure}
\begin{figure} 
\begin{center}     
\includegraphics [height=1.0in, width=4.5in, angle=0]
                 {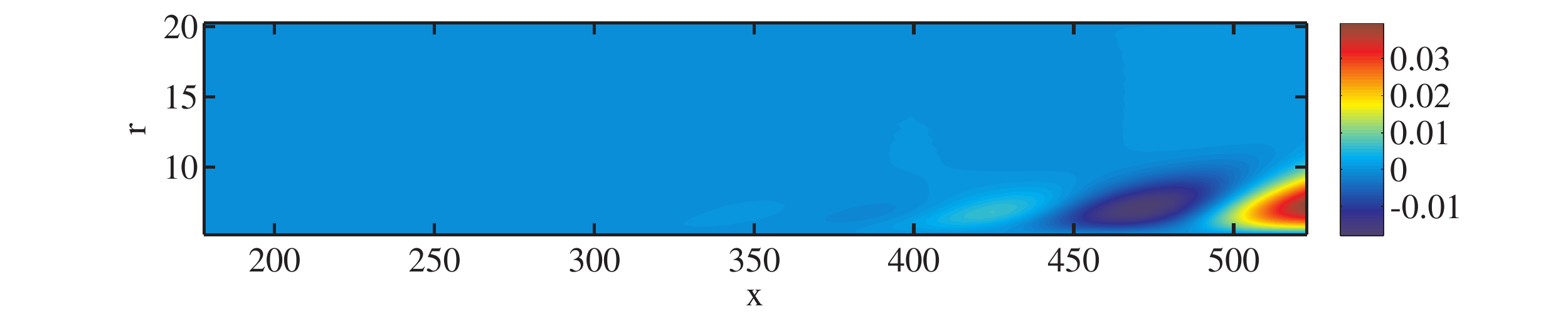}   
\includegraphics [height=1.0in, width=4.5in, angle=0]
                 {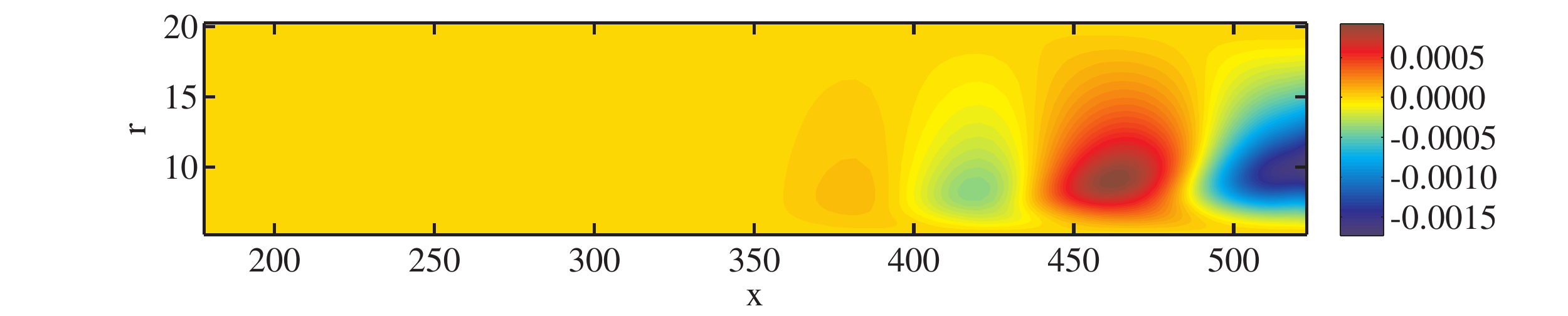}
\includegraphics [height=1.0in, width=4.5in, angle=0]
                 {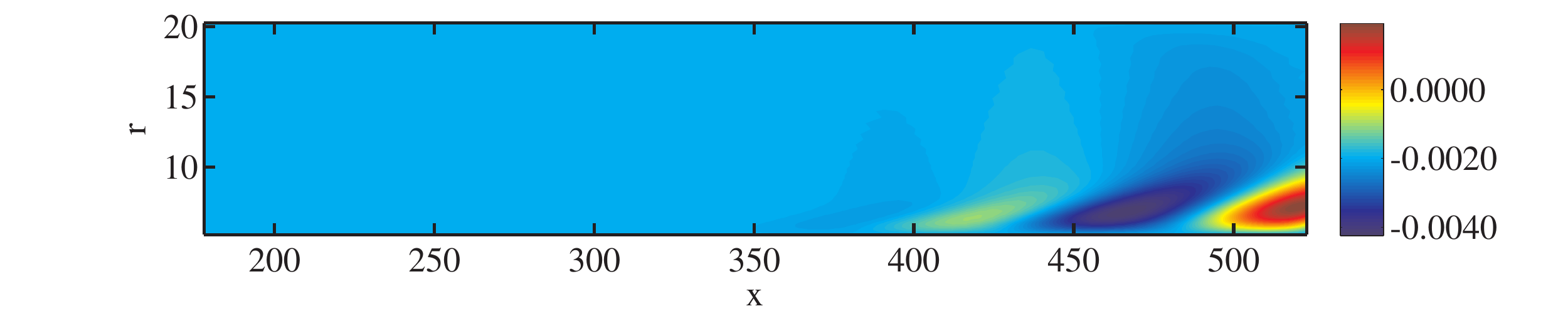}
\caption {Contour plot of (a) streamwise(u) (b) radial (v) and (c) azimuthal (w),
         for disturbance velocity for oscillatory mode  N=1 and Re=383.
         The associated eigenvlaues is $\omega=0.03654-0.01763i$, as marked by 
         square in figure \ref{sp_1_383}.} 
\label{cont_1_383_o}
\end{center}   
\end{figure}

  The eigenspectrum for azimuthal wave number N=1 and $R_{e}=383 $ 
  is shown in figure \ref{sp_1_383}.
  Stationary eigen mode is also found corresponding  to $\omega_r=0$ 
  in the non-axisymmetric case too. 
  The corresponding growth rate $\omega_i$ is -0.02661. 
  Figure \ref{cont_1_383_o} presents the global structure of disturbance amplitude 
  of the oscillatory mode. 
  The oscillatory mode with largest imaginary part is $\omega=0.03654-0.01763i$. 
  This global mode is also temporally stable because $\omega_i < 0$ .
  The eigenmode corresponding to N=1 is non-axisymmetric and therefore, 
  the disturbance velocity have azimuthal component ($w$) also. 
  The spatial structure for streamwise(u), radial (v) and azimuthal(w) 
  disturbance velocity amplitudes is shown in figure \ref{cont_1_383_o}.
  The disturbance amplitudes grows in streamwise direction when moving 
  towards the downstream. 
  The magnitudes of the amplitudes also increases towards the downstream.
  The wave-like nature of the disturbance amplitudes is also found in 
  figure \ref{cont_1_383_o}.
\begin{figure} 
\begin{center} 
\includegraphics [height=1.25in, width=2.0in,angle=0]
                 {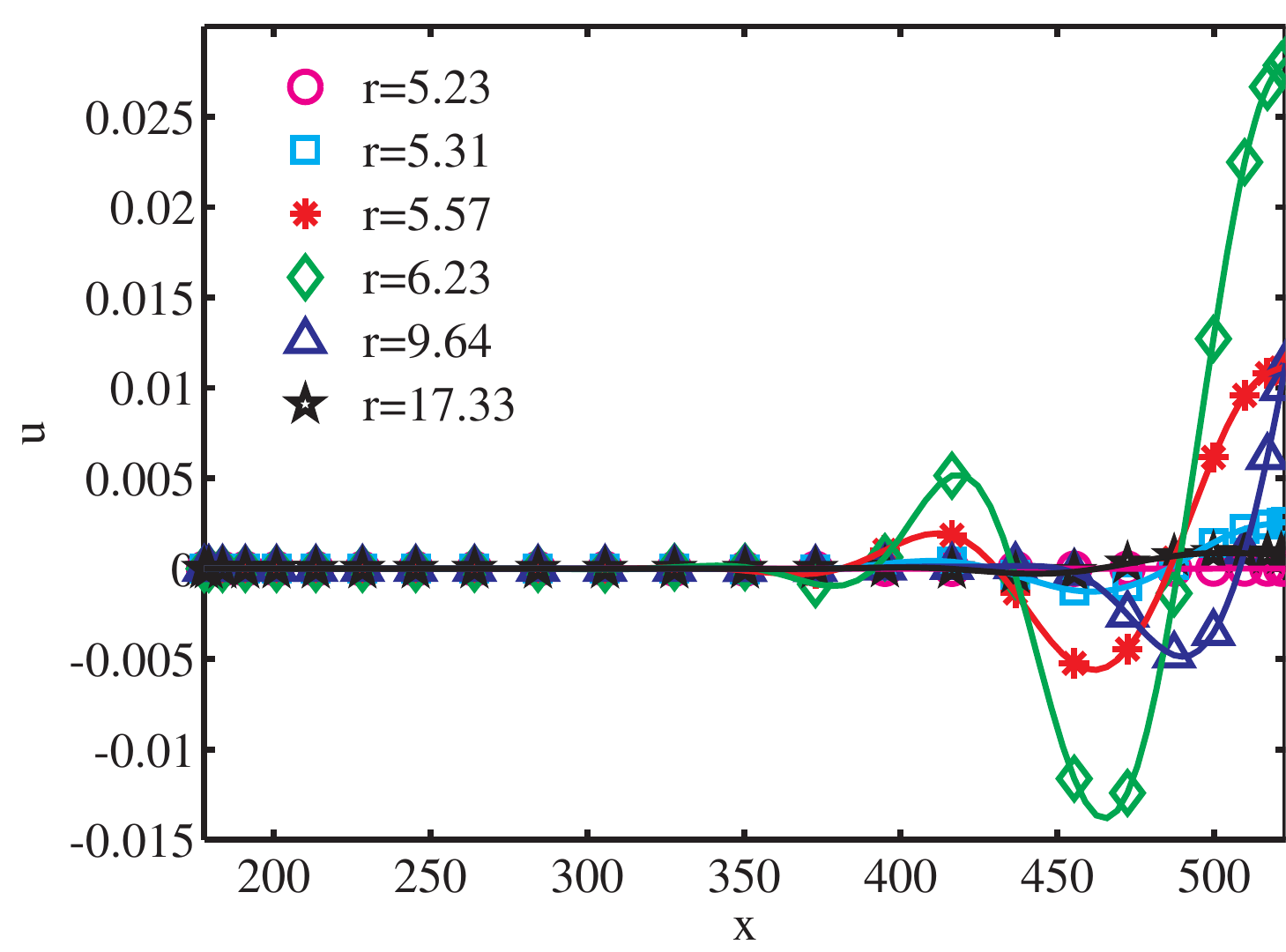} 
\hspace{2px}  
\includegraphics [height=1.25in, width=2.0in, angle=0]
                 {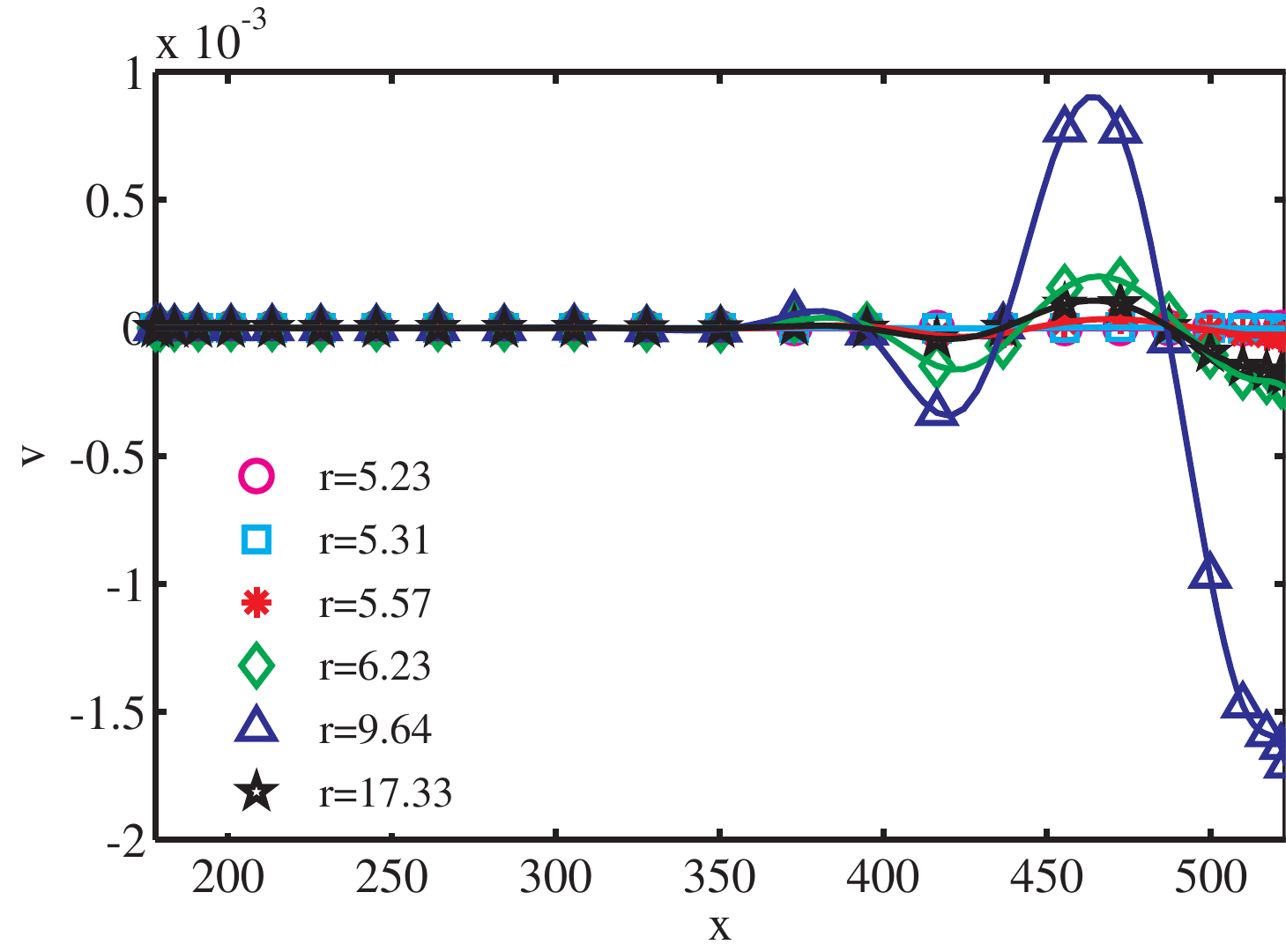} 
\centerline{\includegraphics [height=1.25in, width=2.0in, angle=0]
                             {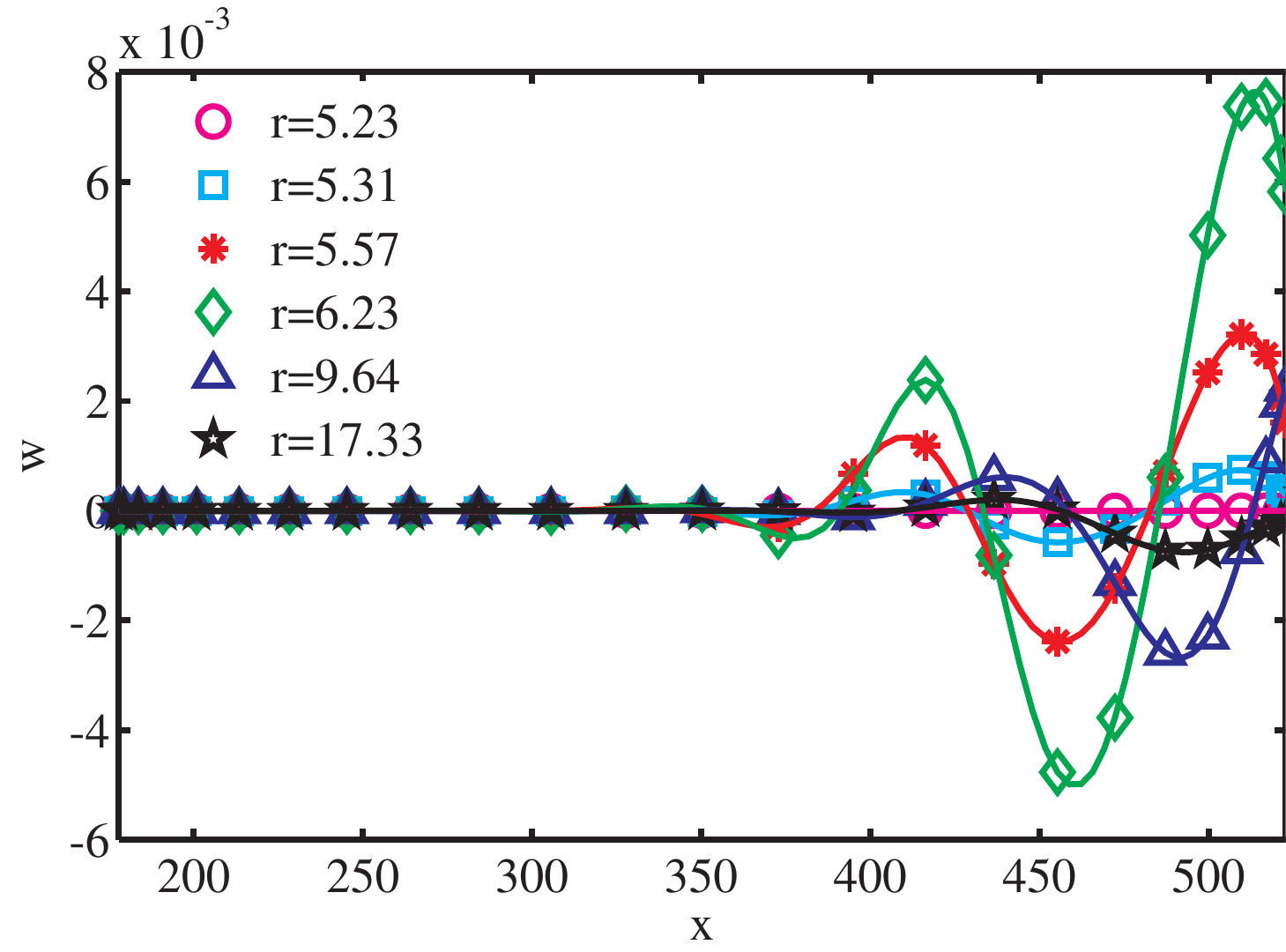}}   
\caption{Variation of disturbance amplitudes in streamwise direction for 
          (a) streamwise (u) (b) radial (v) and (c) azimuthal(w) disturbance 
          amplitudes at different radial location $\omega=0.03654-0.01763i$, 
          as marked by square in figure \ref{sp_1_383}.} 
\end{center}
\label{plot_1_383_o}
\end{figure}
%
%
 \begin{figure}
 \centerline{\includegraphics[height=1.25in, width=2.2in, angle=0] 
                             {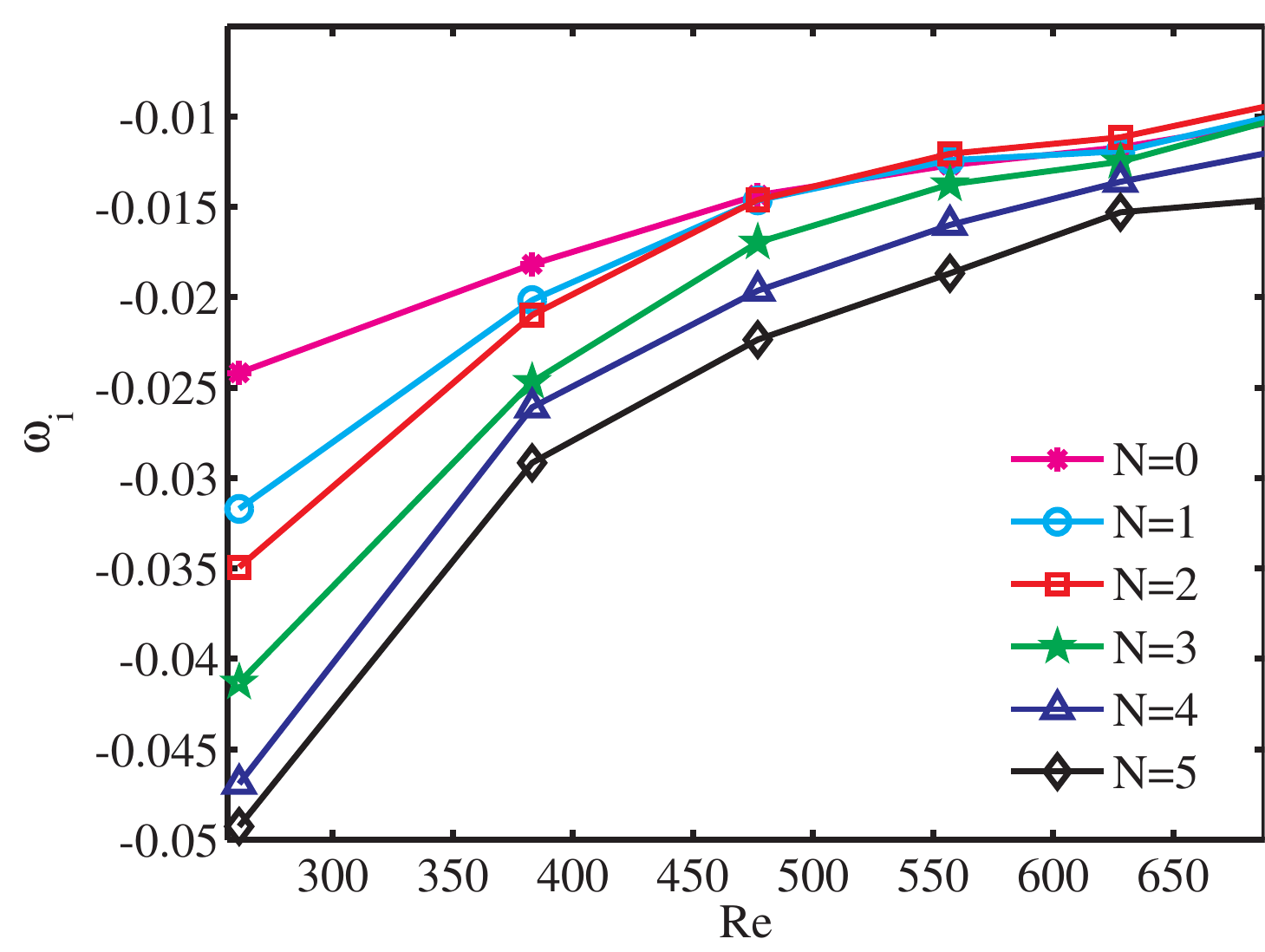}}          
 \caption{Variation in temporal growth rate ($\omega_{i}$) with 
          Reynolds number for different azimuthal wave numbers .}
 \label{temp_growth}
 \end{figure}
 \subsection{Temporal growth rate}
Figure \ref{temp_growth} shows the temporal growth rate of the eigenmodes
for different Reynolds number.
The growth rate increases with the increase in Reynolds number for all 
the azimuthal wave numbers.
As the largest imaginary part is negative for different Reynolds number 
the flow is temporally stable.
The transverse curvature varies inversely with the Reynolds number.
At low Reynolds number growth rate is small and with the increase in 
Reynolds number growth rate increases.
Which proves that transverse curvature has significant damping effect 
on the global temporal mode also.
A global modes with higher wave numbers, N=3,4 and 5 are more stable
than that of axisymmetric and helical mode N=1 and 2.
At low Reynolds number axisymmetric mode is having higher growth rate 
then that of N=1 and 2.
As Reynolds number increases,the growth rate of helical mode N=1 and 2 increases 
then that of axisymmetric mode.
\subsection{Spatial amplification rate}
\begin{figure}  
\begin{center}    
\includegraphics [height=1.25in, width=2.1in, angle=0]{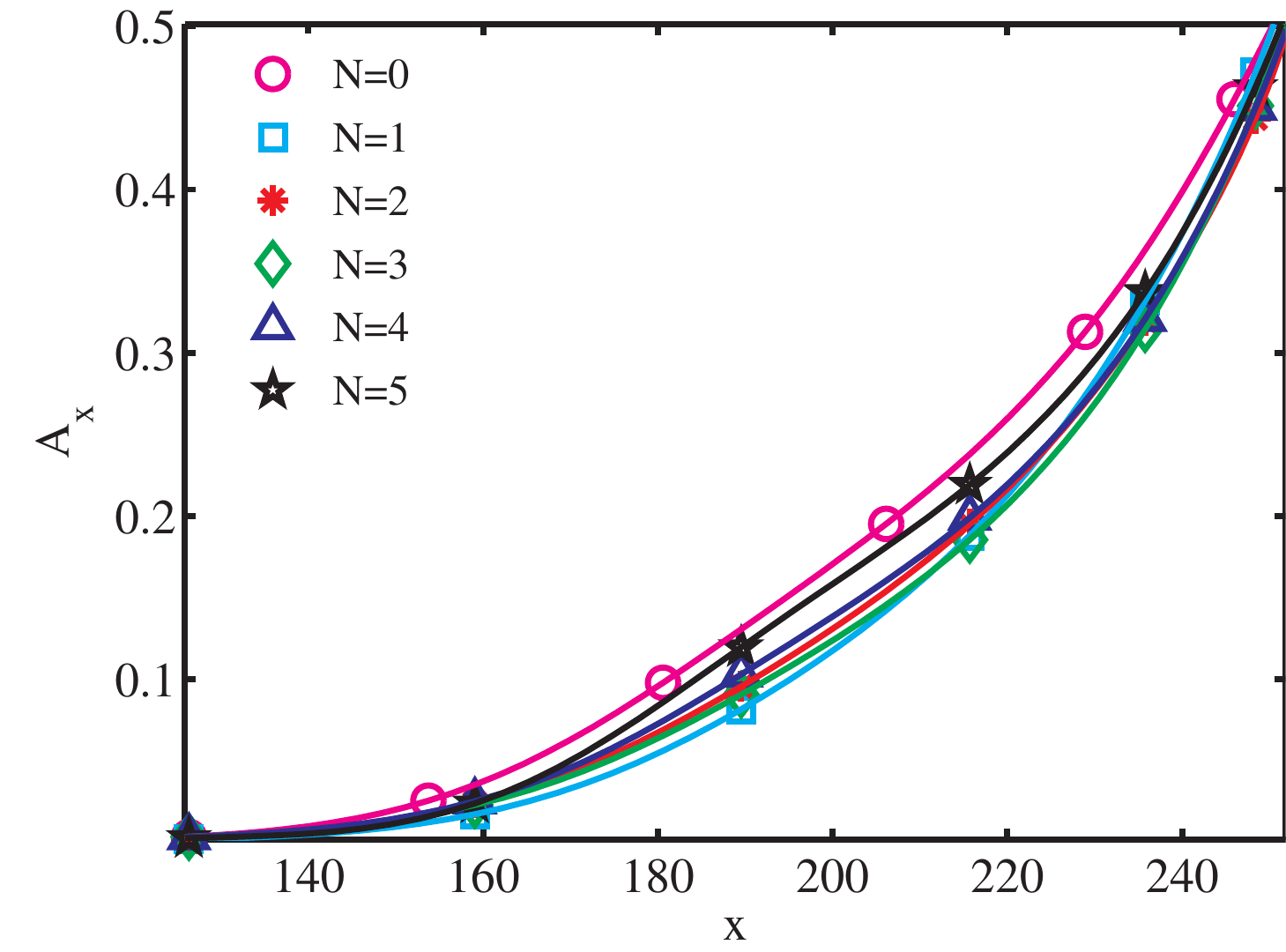}   
\includegraphics [height=1.25in, width=2.1in, angle=0]{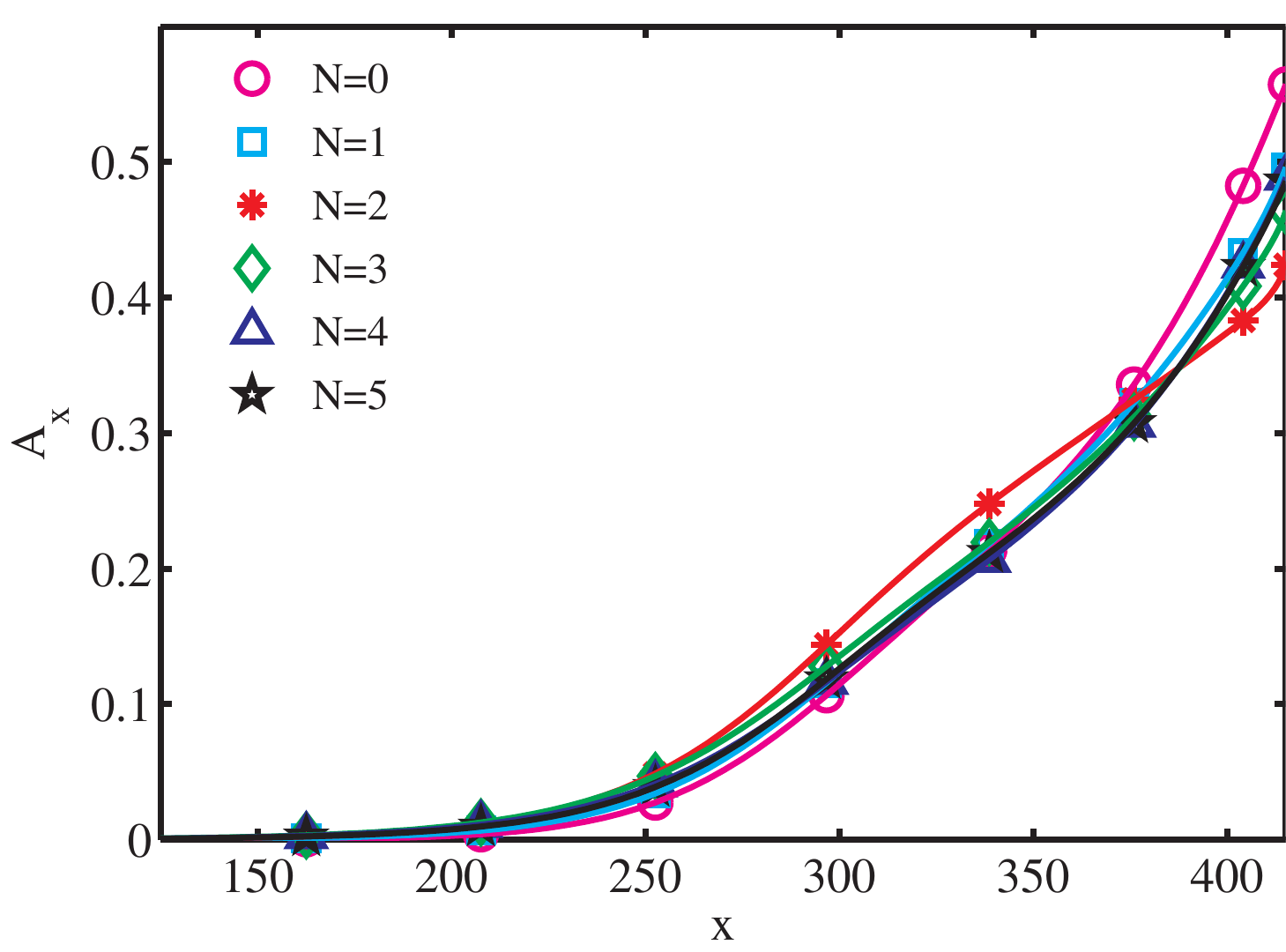}     
\centerline{\includegraphics [height=1.25in, width=2.1in, angle=0]
                             {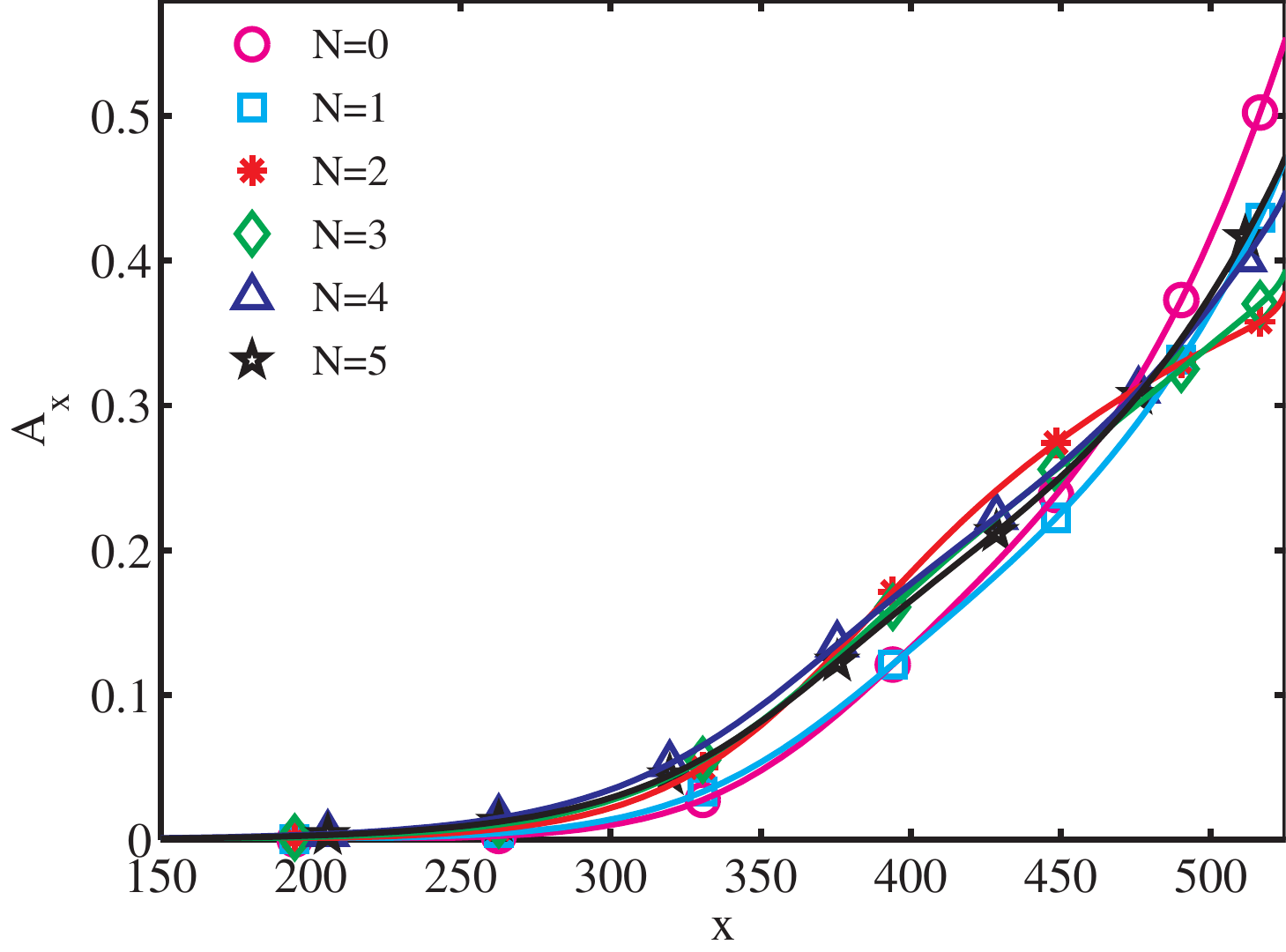}}   
\caption {Spatial amplification rate $A_{x}$ in streamwise direction x for 
         (a) Re=261 (b) Re=477 and (c) Re=628 for different  azimuthal wave 
         numbers. The most unstable temporal mode is considered to compute 
         the spatial amplitude growth $A_x$.} 
\end{center}
\label{sp_amp}
\end{figure}

The global temporal modes exhibit spatial growth/decay in streamwise direction when 
moving towards the downstream.
The disturbances in the streamwise direction at a particular radial location 
may decay or amplify.
However the overall effect of all disturbances together at each streamwise location 
can be quantified by computing the spatial amplitude growth, A(x) \cite {Ehrenstein}.
\begin{equation}
A(x)= \sqrt{\int_{1}^{r_{max}}(u^*(x, r)u(x, r)+v^*(x, r)v(x, r)+w^*(x, r)w(x, r)) dr}
\end{equation}  
where $^*$ denotes the complex conjugate.
Figure \ref{sp_amp} shows the  spatial amplitude growth of the 
disturbance waves for different azimuthal wave numbers at various Reynolds number. 
Most unstable oscillatory ($\omega_r \ne 0$) temporal modes are considered to compute 
the spatial growth of the amplitudes. 
The amplitude growth increases in the streamwise direction as the flow 
is convectively unstable.
At low Reynolds number the the growth of disturbances start at early stage.  
\section{Summary}
The global temporal modes are computed using linear stability theory for 
axisymmetric boundary layer.
The numerical eigenvalue problem is solved using ARPACK, which employs 
Arnoldi's iterative algorithm. 
The stability analysis conducted for various Reynolds number and azimuthal wave numbers.
The largest imaginary part of the computed eigenvalues are negative for the range 
of Reynolds number and azimuthal wave numbers considered here.
Thus the flow is temporally stable.
The global modes in axisymmetric boundary layer, showed wavelike behaviour for the 
range of Reynolds numbers considered. 
The spatial properties of the global mode shows that the disturbances grow in size 
and magnitude within the flow domain in the streamwise direction when move towards 
the downstream, which hints that the flow is convectively unstable.
At low Reynolds number axisymmetric mode (N=0) is least stable one, while at higher Reynolds 
number($Re > 577$) helical mode N=2 is observed least stable one.
The global modes with higher wave numbers N=3, 4 and 5 are more stable.
The comparison of global modes of axisymmetric and flat plate boundary layer 
shows that at low Reynolds number the effect of transverse curvature is significant,
however with the increase in Reynolds number this effect  reduces and global modes 
of the axisymmetric boundary layer become less stable.
Thus, the transverse curvature shows overall damping effect on the global modes. 
The amplitude functions  corresponding to the stationary mode is found to be 
monotonically growing  in the streamwise direction. 
\bibliographystyle{spmpsci}      

%
%
\end{document}